\documentclass[aps,prl,twocolumn,preprintnumbers,amsmath,amssymb,superscriptaddress]{revtex4-2}
\usepackage{amsmath,graphicx}
\usepackage[utf8]{inputenc}
\usepackage[T1]{fontenc}
\usepackage{xcolor}
\usepackage{textcomp}
\usepackage{bm}

\usepackage{siunitx}
\usepackage{physics}
\usepackage{amsmath}
\usepackage{tikz}
\usepackage{mathdots}
\usepackage{yhmath}
\usepackage{cancel}
\usepackage{color}
\usepackage{array}
\usepackage{multirow}
\usepackage{amssymb}
\usepackage{gensymb}
\usepackage{tabularx}
\usepackage{extarrows}
\usepackage{booktabs}
\usetikzlibrary{fadings}
\usetikzlibrary{patterns}
\usetikzlibrary{shadows.blur}
\usetikzlibrary{shapes}

\usepackage{color}
\definecolor{LinkColor}{rgb}{0.75,0.0,0.2}

\usepackage{hyperref}
\hypersetup{
	pdfauthor={good guys},
	pdftitle={good title},
	colorlinks=true,
	citecolor=LinkColor,
	linkcolor=LinkColor,
	urlcolor=LinkColor,
}

\usepackage{listings}
\definecolor{lightgray}{gray}{1}

\usepackage{theorem}

\usepackage{tikz}

\newenvironment{mytikz2}{\begin{tikzpicture}[x=0.4pt,y=0.4pt,yscale=-1,xscale=1,baseline={([yshift=+0ex]current bounding box.center)}]}{\end{tikzpicture}}

\newenvironment{mytikz4}{\begin{tikzpicture}[x=0.6pt,y=0.6pt,yscale=-1,xscale=1,baseline={([yshift=+0ex]current bounding box.center)}]}{\end{tikzpicture}}

\newcommand{\nc}{\newcommand}
\nc{\braoprket}[3]{\langle#1|#2|#3\rangle}
\nc{\opn}[1]{\operatorname{#1}}
\nc{\avg}[1]{\langle#1\rangle}
\nc{\ketbrasame}[1]{|#1\rangle\!\langle#1|}
\nc{\swap}{\opn{SWAP}}
\nc{\E}{\mathbb{E}}
\nc{\Var}{\opn{Var}}
\nc{\dg}{\dagger}

\usepackage[normalem]{ulem}
\nc{\hknew}[1]{\textcolor{brown}{#1}}

\begin{document}
\title{Symmetry restoration and quantum Mpemba effect in symmetric random circuits}

\author{Shuo Liu}
\thanks{These two authors contributed equally to this work.}
\affiliation{Institute for Advanced Study, Tsinghua University, Beijing 100084, China}

\author{Hao-Kai Zhang}
\thanks{These two authors contributed equally to this work.}
\affiliation{Institute for Advanced Study, Tsinghua University, Beijing 100084, China}

\author{Shuai Yin}
\affiliation{School of Physics, Sun Yat-sen University, Guangzhou 510275, China}

\author{Shi-Xin Zhang}
\email{shixinzhang@iphy.ac.cn}
\affiliation{Institute of Physics, Chinese Academy of Sciences, Beijing 100190, China}

\date{\today}

\begin{abstract}
Entanglement asymmetry, which serves as a diagnostic tool for symmetry breaking and a proxy for thermalization, has recently been proposed and studied in the context of symmetry restoration for quantum many-body systems undergoing a quench. In this Letter, we investigate symmetry restoration in various symmetric random quantum circuits, particularly focusing on the U(1) symmetry case. In contrast to non-symmetric random circuits where the U(1) symmetry of a small subsystem can always be restored at late times, we reveal that symmetry restoration can fail in U(1)-symmetric circuits for certain weak symmetry-broken initial states in finite-size systems. In the early-time dynamics, we observe an intriguing quantum Mpemba effect implying that symmetry is restored faster when the initial state is more asymmetric. Furthermore, we also investigate the entanglement asymmetry dynamics for SU(2) and $Z_{2}$ symmetric circuits and identify the presence and absence of the quantum Mpemba effect for the corresponding symmetries, respectively. A unified understanding of these results is provided through the lens of quantum thermalization with conserved charges.
\end{abstract}

\maketitle

\textit{Introduction.---} 
Quantum thermalization is an important topic of fundamental interest. The thermalized state of a chaotic quantum many-body system is linked to the unitary evolution by the eigenstate thermalization hypothesis (ETH)~\cite{PhysRevA.43.2046, PhysRevE.50.888, d2016quantum, rigolThermalizationItsMechanism2008a, Deutsch_2018}. Specifically, the reduced density matrix of a small subsystem $A$ equilibrates to a canonical ensemble: $\rho_{A} \propto e^{-\beta \hat{H}_{A}}$ where $\hat{H}_{A}$ is the system-of-interest Hamiltonian. Furthermore, if $\hat{H}_{A}$ respects U(1) symmetry, the subsystem equilibrates to a grand canonical ensemble: $\rho_{A} \propto e^{-\beta \hat{H}_{A} - \mu \hat{Q}_{A}}$ with $\mu$ denoting the chemical potential and $\hat{Q}_{A}$ representing the conserved charge operator. Consequently, $[\hat{Q}_{A}, \rho_{A}]=0$, and the weak symmetry (also known as average symmetry) for subsystem $A$ can always be restored even with a U(1) symmetry-broken initial state. In other words, symmetry restoration for the small subsystem under quench is an indicator of quantum thermalization. This relation also holds for non-Abelian symmetries with non-commuting conserved charges~\footnote{See Supplemental Material for more details, including (I) symmetry properties of thermal equilibrium states, (II) more numerical results for U(1)-symmetric quantum circuits, (III) numerical results for other symmetric quantum circuits, (IV) analytical results, and Refs.~\cite{YungerHalpern2016_z, Guryanova2016_z, Lostaglio2017_z, YungerHalpern2020_z, Murthy2023_z, Majidy2023a_z, PhysRev.108.171, BALIAN198797, PhysRevLett.103.100403, PhysRevLett.98.050405, doi:10.1126/science.1257026,PhysRevLett.17.1133, colemanThereAreNo1973, PhysRev.158.383,doi:10.1126/science.aau0818,PhysRevLett.121.150501,PhysRevA.100.022324, Collins2006_k}.}. 

In addition to the late-time or equilibrium behaviors, non-equilibrium dynamics have attracted significant attention due to rich interesting phenomena. One example is the counterintuitive Mpemba effect~\cite{EBMpemba_1969} which states that hot water freezes faster than cold water and has been extended in various systems~\cite{PhysRevLett.119.148001, doi:10.1073/pnas.1701264114, PhysRevX.9.021060, kumarExponentiallyFasterCooling2020, bechhoeferFreshUnderstandingMpemba2021, PhysRevLett.131.017101, doi:10.1073/pnas.2118484119, malhotra2024doublempembaeffectcooling}. Quantum versions of the Mpemba effect have also been extensively investigated~\cite{PhysRevB.100.125102, PhysRevLett.127.060401, PhysRevResearch.3.043108, PhysRevA.106.012207, PhysRevE.108.014130, PhysRevLett.131.080402, PhysRevA.110.022213, wang2024mpemba,  PhysRevLett.133.010403} where an external reservoir driving the system out of equilibrium is necessary for the emergence of Mpemba effects. Recently, an intriguing anomalous relaxation phenomenon has been observed in isolated quantum integrable systems~\cite{aresEntanglementAsymmetryProbe2023}. The U(1) symmetry-broken initial states are evolved with the U(1) symmetric Hamiltonian and the weak U(1) symmetry restoration for subsystem $A$ is observed when the subsystem size $\vert A \vert$ is less than half of the total system size $N$~\cite{PhysRevB.89.125101, RevModPhys.83.863, calabreseIntroductionQuantumIntegrability2016, Vidmar_2016,Essler_2016, 10.21468/SciPostPhysLectNotes.18, bastianello2022introduction, alba2021generalized, fagotti2014conservation, bertini2015pre, 10.21468/SciPostPhys.15.3.089}. More importantly, symmetry restoration occurs more rapidly for more asymmetric initial states. This phenomenon is dubbed as the quantum Mpemba effect (QME)~\cite{aresEntanglementAsymmetryProbe2023} and has been demonstrated experimentally on quantum simulation platforms~\cite{PhysRevLett.133.010402}. However, a comprehensive investigation of symmetry restoration and QME in generic chaotic systems is lacking.

Previous work has demonstrated that the U(1) symmetry of $\rho_{A}$ with $\vert A \vert < N/2$ can be restored when the whole system is the random Haar state~\cite{ares2023entanglement},  which can be regarded as the output state of random Haar circuit without U(1) symmetry at late times. This raises a natural question regarding the existence of the QME in the dynamics of random Haar circuits with and without the corresponding symmetry. Moreover, previous investigations on QME have primarily focused on the U(1) symmetry restoration in integrable Hamiltonian dynamics~\cite{aresEntanglementAsymmetryProbe2023, Murciano_2024, PhysRevLett.133.010402, PhysRevLett.133.010401} where the theoretical explanations of QME have hinged on integrability~\cite{PhysRevLett.133.010401}. Although the non-equilibrium dynamics after a global $Z_{2}$ symmetric quantum quench has been investigated before~\cite{Ferro_2024}, no QME has been observed. Therefore, it remains open questions whether QME manifests in the dynamics for other alternative symmetry restoration and whether symmetric random quantum circuits~\cite{doi:10.1146/annurev-conmatphys-031720-030658, PhysRevX.8.021014, PhysRevB.101.104301, PhysRevB.99.174205, PhysRevX.7.031016, PhysRevB.107.L201113, PhysRevLett.132.240402, PhysRevB.110.064323, PhysRevX.9.031009, PhysRevB.98.205136, PhysRevB.100.134306, PhysRevX.12.041002, PhysRevB.107.014308, PhysRevLett.129.120604, PhysRevLett.131.210402, PhysRevB.108.054307, PhysRevLett.131.020401}, play a similar role as the quenched Hamiltonian dynamics. Last but not least, a unified theoretical understanding of symmetry restoration, the QME, and its relation with thermalization is still elusive.

In this Letter, we investigate the dynamics of subsystem symmetry restoration across a range of symmetric and non-symmetric quantum random circuits, considering different initial states. To quantify the degree of symmetry breaking in subsystem $A$, we employ the concept of entanglement asymmetry (EA)~\cite{aresEntanglementAsymmetryProbe2023, 10.21468/SciPostPhys.15.3.089, Ferro_2024, Marvian2014_z}, which has been extensively studied as an effective symmetry broken measure in out-of-equilibrium many-body systems~\cite{PhysRevLett.133.010401, Khor2024confinementkink} and quantum field theories~\cite{capizziEntanglementAsymmetryOrdered2023, capizzi2023universal, PhysRevD.109.065009}. It is defined as 
\begin{eqnarray}
    \Delta S_{A}  = S(\rho_{A, Q})   - S(\rho_{A}).
\end{eqnarray}
Here $S(\rho_{A})$ represents the von Neumann entropy of subsystem $\rho_{A}$, and $\rho_{A, Q} = \sum_{q} \Pi_{q} \rho_{A} \Pi_{q}$ where $\Pi_{q}$ is the projector to the $q$-th eigensector of the corresponding symmetry operator $g$. In the case of U(1) symmetry, the symmetry operator $g$ is the total charge operator in subsystem $A$, $\hat{Q}_{A} = \sum_{i}^{\vert A \vert} \sigma_{i}^{z}$ and the computational basis coincides with the eigenbasis of $\hat{Q}_{A}$. We also extend the definition of EA to SU(2) cases for the first time, where a carefully designed unitary transformation is required for $\rho_{A}$ to properly address non-commuting conserved charges~\cite{Note1}. It is worth noting that $\Delta S_{A} \geq 0$ by definition and it only vanishes when $\rho_{A}$ is block diagonal in the eigenbasis of the symmetry operator.
Symmetry restoration as indicated by $\Delta S_{A}=0$ is also a necessary condition for quantum thermalization due to the thermal equilibrium form of the mixed state. Due to the randomness in circuit configurations, we focus on the average EA, 
$\mathbb{E}[\Delta S_{A}]$. In the theoretical analysis, we utilize R\'enyi-2 EA, $\mathbb{E} [\Delta S^{(2)}_{A}]$, by replacing von Neumann entropy with R\'enyi-2 entropy for simplicity, which shows qualitatively the same behaviors as EA.

Based on the rigorous theoretical analysis and extensive numerical simulations, we have revealed that the subsystem symmetry restoration of the U(1)-symmetric circuits depends on initial states, which significantly differs from the case of random Haar circuits where the dynamics are agnostic of different initial states. Specifically, when starting from a tilted ferromagnetic state with a sufficiently small tilt angle $\theta$, we demonstrate that the late-time EA remains non-zero in finite-size systems, indicating that the final state remains symmetry broken. The persistent symmetry breaking in finite-size systems is a universal feature for U(1) symmetric dynamics~\cite{Note1}. Conversely, the symmetry can always be restored when the initial state is more U(1)-asymmetric with a large tilt angle as long as $\vert A \vert /N <1/2$. 

More importantly, we have also observed the emergence of QME in the EA dynamics of U(1)-symmetric circuits, which is absent in random Haar circuits without symmetry. The emergence of QME in chaotic systems can be understood through the lens of quantum thermalization. The thermalization speed varies significantly across different charge sectors~\cite{Note1} (see also results for thermalization in U(1)-symmetric circuits~\cite{chang2024deepthermalizationchargeconservingquantum, li2024efficientquantumpseudorandomnessconservation}). Specifically, charge sectors with small Hilbert space dimensions do not obey ETH and thus do not thermalize or thermalize slowly.
Consequently, symmetry restoration is slow when the subsystem of the initial state has a large overlap with the charge sector of a small dimension. Therefore, a QME occurs when the state with a smaller initial EA simultaneously has a larger overlap with the small charge sector resulting in slow thermalization, as in the case of the tilted ferromagnetic state. In sum, the mechanism behind QME is attributed to slower thermalization induced by more symmetric initial states. We have validated this theoretical understanding with different initial states and internal symmetries.

\textit{Setup.---} For the U(1) symmetry restoration, inspired by the previous studies~\cite{aresEntanglementAsymmetryProbe2023}, we adopt a tilted ferromagnetic state as the initial state (see more numerical results with different initial states in the SM~\cite{Note1}) defined as 
\begin{eqnarray}
    \vert \psi_{0} (\theta)\rangle =  e^{-i\frac{\theta}{2} \sum_{j} \sigma_{j}^{y}} \vert 000...0 \rangle,
    \label{eq: tilted ferromagnetic}
\end{eqnarray}
where $\sigma_{j}^{y}$ is the Pauli-Y operator on $j$-th qubit and the tilt angle $\theta$ determines the charge asymmetric level of the initial states: when $\theta=0$, $\vert \psi_{0} (0) \rangle= \vert 000...0\rangle$ is U(1)-symmetric and $\Delta S_{A}=0$ for any subsystem $A$. As $\theta$ increases, $\Delta S_{A}$ also increases until it reaches its maximal value at $\theta=\pi/2$. 

\begin{figure}
\centering
\includegraphics[width=0.38\textwidth, keepaspectratio]{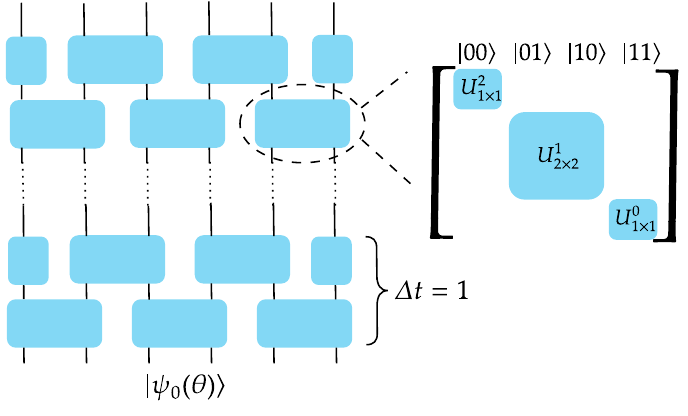}
\caption{Random circuit with $6$ qubits. The initial state is chosen as the tilted ferromagnetic state and the blue rectangle represents random two-qubit gates in the even-odd brick-wall pattern. For U(1)-symmetric circuits, each two-qubit gate respects U(1) symmetry, resulting in a block diagonal structure for the unitary matrix of quantum gates.}
\label{fig:setup}
\end{figure}

As shown in Fig.~\ref{fig:setup}, the initial state undergoes the unitary evolution of random quantum circuits with periodic boundary conditions where two-qubit gates are arranged in a brick-wall structure. In the case of non-symmetric circuits, each two-qubit gate is randomly chosen from the Haar measure. For the U(1)-symmetric case, the matrix for each two-qubit gate is block diagonal as shown in Fig.~\ref{fig:setup} and each block is randomly sampled from the Haar measure. One discrete time step $\Delta t=1$ includes two layers of two-qubit gates. We calculate the EA dynamics $\mathbb{E} [\Delta S_{A}] $ of subsystem $A$ averaged over different circuit configurations to monitor the dynamical and steady behaviors of symmetry restoration.

\textit{Symmetry restoration in the long-time limit.---} We approximate the long-time limit of random circuit ensemble with a simpler ensemble $\mathbb{U}$ for a single random unitary $U$ acting on all qubits to compute R\'enyi-2 EA~\cite{Note1, Brandao2016_k, hearth2023unitary, li2023designs, li2023d}. For the non-symmetric random circuit evolution, $\mathbb{U}$ constitutes a global $2$-design for the Haar measure. Consequently, the average R\'enyi-2 EA at late time is~\cite{ares2023entanglement}
\begin{equation}\label{eq:non-symmetric}
    \mathbb{E} [ \Delta S_A^{(2)} ] \approx -\log\left[ \frac{1 + 2^{2|A|-N}/\sqrt{\pi |A|} }{1 + 2^{2|A|-N}} \right].
\end{equation} 
With large $N$,
$\mathbb{E} [\Delta S_A^{(2)} ]$ approches zero if $|A|<N/2$ while it sharply changes to a non-zero value $\log\sqrt{\pi|A|}$ if $|A|>N/2$. Therefore, the broken symmetry of subsystem $A$ with $|A|<N/2$ can always be restored by the non-symmetric random circuits at a late time, regardless of the initial states.

\begin{figure}
\centering
\includegraphics[width=0.98\linewidth, keepaspectratio]{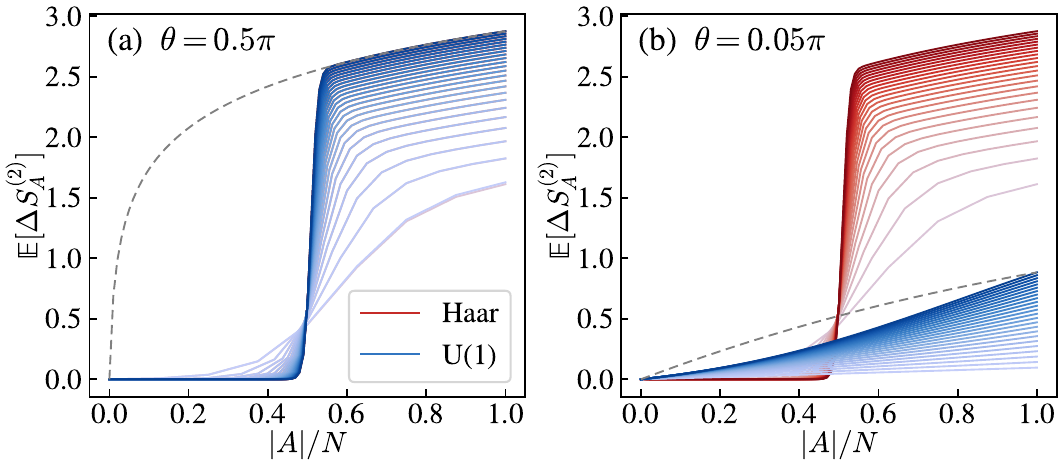}
\caption{The average R\'enyi-2 EA $\mathbb{E} [\Delta S^{(2)}_A ]$ in the long-time limit of random circuit evolution starting from the tilted ferromagnetic initial states with tilt angle (a) $\theta=0.5\pi$ and (b) $\theta=0.05\pi$ versus the subsystem size $|A|$. The increasing intensity of colors represents increasing system sizes $N\in\{8,12,\ldots,100\}$. The blue and red lines represent the results for U(1)-symmetric random circuits and random circuits without symmetry restriction, respectively. The grey dashed lines represent the R\'enyi-2 EA for the initial states with $N=100$.}
\label{fig:deltaS_vs_A}
\end{figure}

For the U(1)-symmetric random circuit evolution, $\mathbb{U}$ is a global $2$-design for the composition of the Haar measures over each charge sector~\cite{hearth2023unitary, li2023designs, li2023d}. The average R\'enyi-2 EA can be accurately obtained by calculating certain summations of polynomial numbers of binomial coefficient products, which arise from counting charge numbers~\cite{Note1}. 
Under the condition of large system size and large tilt angle, a simplified analytical form of $\mathbb{E} [\Delta S_A^{(2)}]$ can be obtained by approximating the binomial coefficients with the Gaussian distributions,
\begin{equation}\label{eq:symmetric}
    \mathbb{E} [\Delta S_A^{(2)} ] \approx -\log \left[ \frac{1 + g(\theta)^{2|A|-N} /\sqrt{\pi|A|} }{  1 + g(\theta)^{2|A|-N}} \right],
\end{equation}
which resembles the non-symmetric case in Eq.~\eqref{eq:non-symmetric} except for the $\theta$-dependent base factor
\begin{equation}
    g(\theta) = 2 \exp\left[ -\frac{1}{2}  \log^2\left(\tan^2\frac{\theta}{2}\right)\right].
\end{equation}
If $\theta=0.5\pi$, Eq.~\eqref{eq:symmetric} coincides with Eq.~\eqref{eq:non-symmetric}, and thus the late-time EA are the same as shown in Fig.~\ref{fig:deltaS_vs_A}(a). If the tilt angle remains large but deviates from $0.5\pi$, Eq.~\eqref{eq:symmetric} indicates that the main characteristic remains unchanged compared with the non-symmetric case: the symmetry is restored for a small subsystem of $|A|<N/2$ but still broken for $|A|>N/2$.

\begin{figure}
\centering
\includegraphics[width=0.98\linewidth, keepaspectratio]{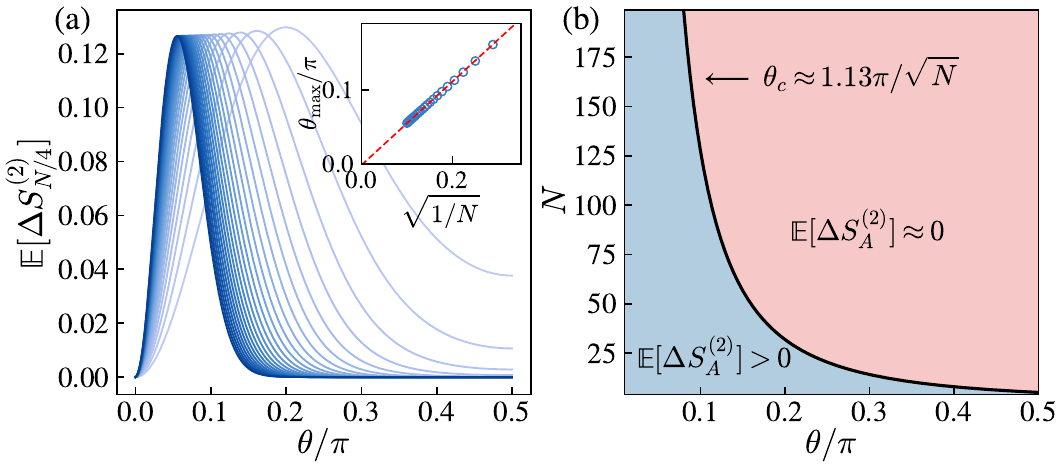}
\caption{(a) The average R\'enyi-2 EA $\mathbb{E} [\Delta S^{(2)}_A]$ in the long-time limit of U(1)-symmetric random circuit evolution at $|A|=N/4$ versus tilt angle $\theta$. The increasing intensity of colors corresponds to $N\in\{8,12,\ldots,100\}$. The inset depicts the peak position $\theta_{\mathrm{max}}$ versus $N$. (b) The crossover for the symmetric restoration at $|A|<N/2$. The red and blue areas represent the symmetry-restored and persistent symmetry-breaking behaviors, respectively. The ``critical value'' $\theta_c\approx 2\theta_{\mathrm{max}}$ for the finite size crossover depends on $N$ with a $1/\sqrt{N}$ scaling.}
\label{fig:deltaS_vs_theta}
\end{figure}

However, if the tilt angle is sufficiently small, the Gaussian approximation fails. We rely on direct estimation of the summations of binomial coefficients~\cite{Note1}, which shows that for small tilt angles such as $\theta<0.1\pi$, $\mathbb{E} [\Delta S_A^{(2)}]$ will converge to a significant finite value in the long-time limit even for $|A|<N/2$ and large finite $N$ as shown in Fig. ~\ref{fig:deltaS_vs_A}(b).
In other words, when the symmetry breaking in the tilted ferromagnetic initial state is relatively weak, it becomes challenging to fully restore the subsystem symmetry through the U(1)-symmetric random circuit evolution. Conversely, those initial states exhibiting more severe symmetry breaking can restore the symmetry successfully instead. This phenomenon is reminiscent of an extreme limit of the QME, where instead of restoring slowly, the symmetry does not fully restore for initial states with weak symmetry breaking. We remark that the persistent symmetry breaking is a universal feature of U(1)-restoring dynamics quenching a sufficiently weak symmetry-breaking tilted ferromagnetic state~\cite{Note1}. As shown in Fig.~\ref{fig:deltaS_vs_theta}, there exists a critical value $\theta_c$ where on the small-$\theta$ side the symmetry is not restored, leaving a persistent symmetry-breaking behavior. It is worth noting that the above discussions only apply to the finite-size system as the critical value $\theta_{c}$ slowly varies with the system size $N$ with a scaling of $\theta_c \approx 1.13 \pi/\sqrt{N}$.

\textit{Quantum Mpemba effect in early time dynamics.---}  Now we proceed to consider the EA dynamics for different initial states with varying tilt angles $\theta$. The numerical simulations are performed using the TensorCircuit package~\cite{Zhang2023tensorcircuit}. We observe that EA decays more rapidly as the tilt angle increases for U(1) symmetric random quantum circuits, as illustrated in Fig.~\ref{fig:EAdynamics} (a). Namely, a QME emerges in the symmetry restoration process. It is important to highlight that the presence of QME depends on the specific initial states and we observe the absence of QME with initial tilted N\'eel states as shown in Fig.~\ref{fig:EAdynamics_diffinit1} (b). Nevertheless, we emphasize that the presence of QMEs is not a fine-tuned phenomenon. We further investigate the EA dynamics and observe QME with various sets of initial states, including tilted ferromagnetic states where the tilt angle on each qubit is randomly sampled from $[-W, W]$, which is more compatible with the experimental demonstration of QMEs on quantum devices since it doesn't require high precision state preparation~\cite{Note1}.

\begin{figure}
\centering
\includegraphics[width=0.47\textwidth, keepaspectratio]{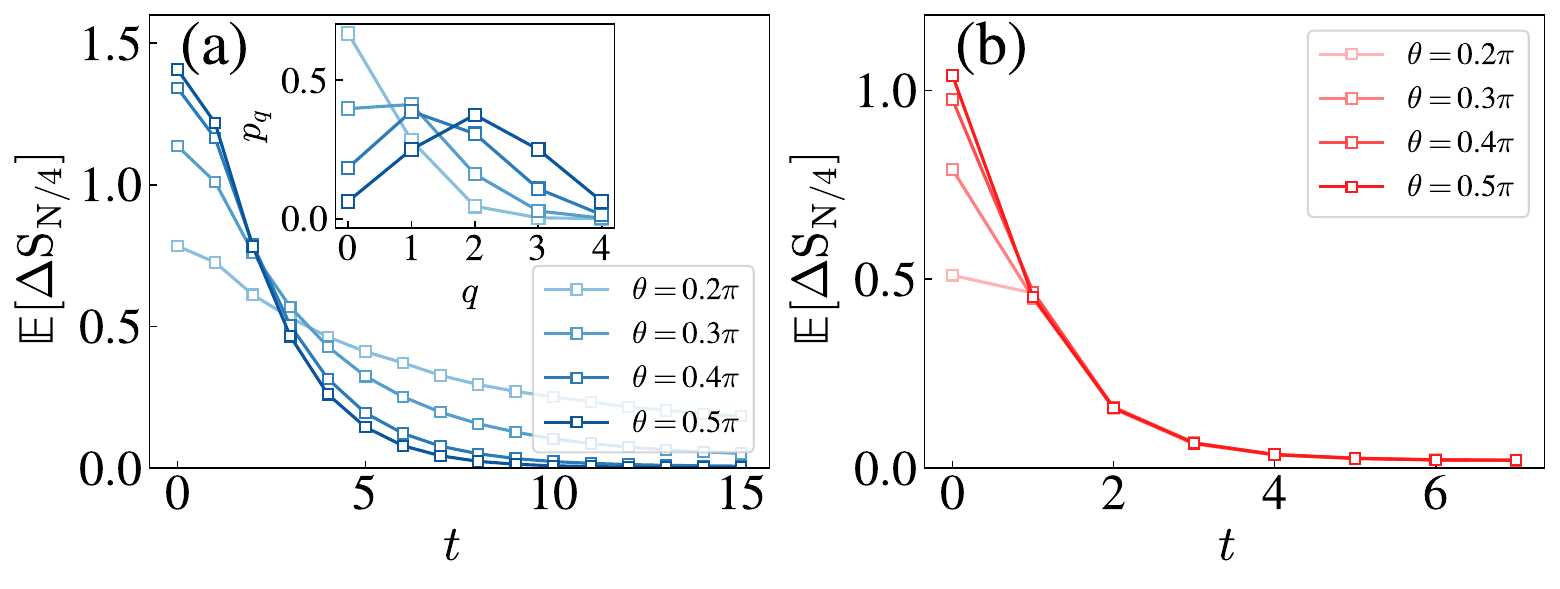}
\caption{(a)(b) show the EA dynamics of subsystem $A=[0, N/4]$ for random quantum circuits with and without U(1) symmetry respectively. We use $N=16$ in the former case and $N=8$ in the latter case. The Inset of (a) shows the overlaps of $\rho_{A}$ with different charge sectors ranging from $\{0, \ldots, N/4\}$.  QME exists for U(1)-symmetric random circuits while it is absent in random circuits without any symmetry.}
\label{fig:EAdynamics}
\end{figure}

On the contrary, for non-symmetric random circuits, EA dynamics with different initial states coincide
as depicted in Fig.~\ref{fig:EAdynamics} (b). Consequently, although the EA still tends to zero at late times, the QME disappears trivially. This behavior can be understood in terms of the effective statistical model, where the initial state dependence has been eliminated as the inner product between different initial product states and the first layer of random unitary gates is constant~\cite{Note1}.

In addition, we have conducted investigations on setups that incorporate additional symmetries, such as spatial or temporal translational symmetry (random Floquet circuit)~\cite{PhysRevB.98.134204, FloquetClifford, PhysRevB.109.024311, PhysRevX.8.041019}. We have found that thermalization accompanied by symmetry restoration and the QME persists in these setups as well~\cite{Note1}. Interestingly, the temporal translation symmetry slows down the symmetry restoration, consistent with the slow thermalization results in \cite{PhysRevB.109.024311}. 

To understand the unified mechanism behind QME in generic chaotic systems, we first consider the overlaps between reduced density matrix $\rho_{A}$ and different charge sectors, defined as $p_{q} = {\rm tr}(\Pi_{q} \rho_{A} \Pi_{q}) $. As shown in the inset of Fig.~\ref{fig:EAdynamics} (a), the charge distribution is more peaked to the charge sector of a small dimension as the decreases of tilt angle $\theta$, i.e., the initial state is more symmetric. Furthermore, the thermalization speeds for different charge sectors are conjectured to be different where charge sectors with $O(1)$ dimension generally fail the ETH and thermalize slowly. The conjecture of thermalization speed dependence on the Hilbert subspace dimension is numerically validated in the SM~\cite{Note1} which is of standalone importance toward a better understanding of quantum thermalization. Therefore, for tilted ferromagnetic states, weaker initial subsystem symmetry breaking is linked with slower thermalization speed via the larger overlap with small charge sectors. Consequently, QME occurs as the symmetry restoration is slower for more symmetric initial states.

\begin{figure}
\centering
\includegraphics[width=0.47\textwidth, keepaspectratio]{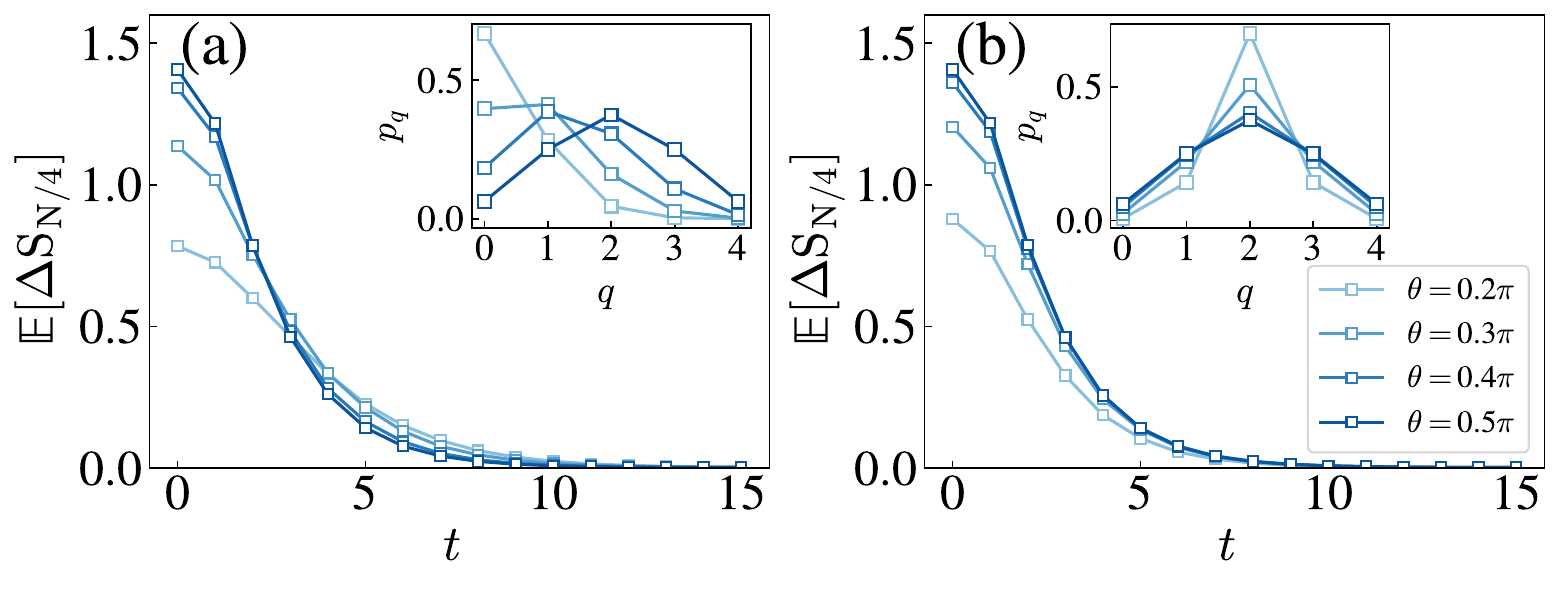}
\caption{EA dynamics with initial (a) tilted ferromagnetic state with a middle domain wall and (b) tilted N\'eel state. Insets show the overlaps of $\rho_{A}$ with different charge sectors $ q \in \{0, \ldots, N/4\}$.
In the former case, the QME is present, whereas it is absent in the latter case, although the late-time behaviors are the same for both initial states based on the analytical results.}
\label{fig:EAdynamics_diffinit1}
\end{figure}

This unified mechanism provides insights to identify the suitable initial states exhibiting QME beyond the tilted ferromagnetic state extensively investigated before. To further validate the thermalization explanation of QME, we investigate the EA dynamics of two different initial states: one is the tilted ferromagnetic state with a middle domain wall and the other is the tilted N\'eel state~\cite{Note1}. The steady-state EA should be the same for these two initial states as discussed in the SM~\cite{Note1}. However, QME is captured by the early-time behaviors of EA dynamics, making the local spin configurations of the initial states crucial due to the casual cone structure. The numerical results of overlaps between the two types of initial states and different charge sectors are shown in Fig.~\ref{fig:EAdynamics_diffinit1}. For the tilted ferromagnetic state with a middle domain wall, the overlap distribution is similar to that of the tilted ferromagnetic state. Therefore, more symmetric states thermalize slower and the QME occurs. On the contrary, for the tilted N\'eel state, its charge distribution is always peaked to the largest charge sector and thus it strongly obeys the ETH
regardless of tilt angles $\theta$. Consequently, the thermalization speed is essentially unchanged and the QME is absent.

Moreover, we also investigate the symmetry restoration dynamics in quantum circuits with SU(2) and $Z_{2}$ symmetries. We extend the definition of EA for SU(2) symmetry and identify the presence (absence) of QME in the SU(2) ($Z_{2}$) symmetric circuits~\cite{Note1}. The unified mechanism above also provides insights into these different internal symmetry cases. For $Z_{2}$ symmetry, there are only two equally large Hilbert subspaces, and the thermalization speeds of different initial states are expected to be similar with no QME. In the SU(2) symmetry case, the dimensions of different symmetry sectors vary from constant to exponential scaling like the U(1) case. Therefore, QME can be observed for carefully designed initial states satisfying the criteria above.

\textit{Conclusions and discussions.---} In this Letter, we have presented a rigorous and comprehensive theoretical analysis of subsystem symmetry restoration under the evolution of random quantum circuits respecting the U(1) symmetry. Our findings reveal that U(1)-symmetric circuits hinder the U(1) symmetry restoration when the input is a tilted ferromagnetic initial state with a small tilt angle $\theta$. Conversely, the symmetry can always be restored when the tilt angle is large, i.e., the initial state is more U(1)-asymmetric. These results highlight the distinctions between U(1)-symmetric and non-symmetric circuits in terms of symmetry restoration and quantum thermalization. 

More importantly, besides the late-time analytical results, we have numerically investigated the early-time dynamics of symmetry restoration and provided a unified understanding of QME in generic chaotic systems in the context of quantum thermalization. We have validated this theoretical understanding via the correct predictions of the presence and absence of QME for various initial states and internal symmetries that have not been explored before.

There are various interesting questions worth further investigation. For example, the comprehensive extension of the symmetry restoration of other internal symmetries in both integrable and chaotic systems. Additionally, the investigation of symmetry restoration and the QME in the many-body localized systems is lacking. Addressing this gap will significantly enhance our theoretical comprehension of the mechanisms underlying the QME across diverse systems~\cite{MBLMpemba}.

\textit{Acknowledgement.---}  We acknowledge helpful discussions with Rui-An Chang, Zhou-Quan Wan, and Han Zheng. This work was supported in part by the Innovation Program for Quantum Science and Technology (grant No. 2021ZD0302502). SL acknowledge the support during the visit to Sun Yat-sen University. S. Yin is supported by the National Natural Science Foundation of China (Grants No. 12075324 and No. 12222515) and the Science and Technology Projects in Guangdong Province (Grants No. 211193863020). The work of S.X.Z. is supported in part by a startup grant at IOP-CAS.

\let\oldaddcontentsline\addcontentsline
\renewcommand{\addcontentsline}[3]{}
%

\clearpage
\newpage
\widetext

\begin{center}
\textbf{\large Supplemental Material for ``Symmetry restoration and quantum Mpemba effect in symmetric random circuits''}
\end{center}

\date{\today}
\maketitle

\renewcommand{\thefigure}{S\arabic{figure}}
\setcounter{figure}{0}
\renewcommand{\theequation}{S\arabic{equation}}
\setcounter{equation}{0}
\renewcommand{\thesection}{\Roman{section}}
\setcounter{section}{0}
\setcounter{secnumdepth}{4}

\addtocontents{toc}{\protect\setcounter{tocdepth}{0}}
{
\tableofcontents
}

\section{Symmetry properties of thermal equilibrium states}
For a nonintegrable system under Hamiltonian $H$ with some conserved commuting or noncommuting charges $Q_a$, the thermal equilibrium state is given by quantum thermodynamics as \cite{YungerHalpern2016_z, Guryanova2016_z, Lostaglio2017_z, YungerHalpern2020_z, Murthy2023_z, Majidy2023a_z}
\begin{equation}
    \rho_{th} = \frac{1}{Z}e^{-\beta (H-\sum_a \mu_a Q_a)},
\end{equation}
where $Z$ is the partition function for normalization $Z=\text{Tr}(e^{-\beta (H-\sum_a \mu_a Q_a)})$, and these coefficients of $\beta$ and $\mu$ are determined by the expectation value $E=\text{Tr}(\rho_{th}H)$ and $\langle Q_a\rangle = \text{Tr}(\rho_{th}Q_a)$. When the charges are noncommuting as for non-Abelian symmetries of $H$, the equilibrium state is often referred as non-Abiliean thermal states (NATS) as the counterpart of generalized Gibbs ensemble (GGE)~\cite{PhysRev.108.171, BALIAN198797, PhysRevLett.103.100403, PhysRevLett.98.050405, doi:10.1126/science.1257026} for integrable systems.

Suppose all charges $Q_a$ are commuting with each other. In that case, we can easily find a unitary transformation $U$ such that $U^\dagger\vert i\rangle= \vert Q_a=Q_a(i)\rangle$ which transforms the computational basis $\vert i\rangle $ into the common eigenstate of multiple charges $Q_a$. For example, for $Q=J_z$, $U=I$ already satisfies the requirement. The transformed density matrix gives
\begin{equation}
    U\rho_{th}U^\dagger = \frac{1}{Z}e^{-\beta (UHU^\dagger-\sum_a \mu_a UQ_a U^\dagger)}.
\end{equation}
The matrix on the exponential after the transformation is block diagonal for different charge sectors $\{Q_a\}$. We can show this by checking the matrix elements as $\langle i\vert  UHU^\dagger\vert j\rangle=\langle Q_a=Q_a(i)\vert H\vert Q_a=Q_a(j)\rangle\propto\delta_{Q_a(i),Q_a(j)}$ and $\langle i\vert UQ_aU^\dagger\vert j\rangle=\langle Q_a=Q_a(i)\vert Q_a\vert Q_a=Q_a(j)\rangle=Q_a(i)\delta_{i,j}$. The matrix exponential of the block diagonal matrix by charge sectors is still a block matrix of the same structure. Therefore, a thermal equilibrium state must admit a block diagonal structure on the basis of commuting charges (or equivalently, after applying the similar transformation $U$). Such a block diagonal structure is a manifestation of the symmetry restoration for corresponding charges. We thus conclude that symmetry restoration is a necessary condition for thermalization. Another way to understand the symmetry restoration in a one-dimensional system is provided by Mermin-Wagner-Hohenberg theorem~\cite{PhysRevLett.17.1133, colemanThereAreNo1973, PhysRev.158.383}. The finite energy density of the initial symmetry-broken state acts as an effective non-zero temperature and leads to general symmetry restoration because the Mermin-Wagner-Hohenberg theorem forbids spontaneous symmetry breaking at a finite temperature in 1D.

Now we focus on the noncommuting charges case. We use SU(2) symmetry as a representative example in the following analysis. Suppose we choose the $J_z$ direction arbitrarily and apply a similar procedure as the Abelian case, namely, we introduce $U^\dagger \vert i\rangle = \vert J, J_z\rangle$, where $J$ and $J_z$ is a set of good quantum numbers characterizing quantum states as the basis of SU(2) group irreducible representations. We now have $\langle i\vert UHU^\dagger\vert j\rangle\propto \delta_{J(i),J(j)}\delta_{ J_z(i),J_z(j)}$ and $\langle i\vert UJ_zU^\dagger\vert j\rangle = J_{z}(i) \delta_{i, j}$. The two terms keep the expected block diagonal structure from SU(2) symmetry. However, $U J_{x, y} U^\dagger$ has non-diagonal non-zero terms between states of different $J_z$, thus leaving the final density matrix with no explicit block diagonal structure.

To overcome the above issue, we need to find a better, more fine-grained transformation $U$, such that $U^\dagger \vert i\rangle = \vert J, J_z'\rangle$. And now the $J_z'$ direction is properly determined, such that we have $\text{Tr}(\rho_{th}J_{x'})=\text{Tr}(\rho_{th}J_{y'})=0$. We can decompose the new transformation $U$ as $U_0R$.  $U_0$ is the same as the above paragraph, composed of CG coefficients, transforming the computational basis into the SU(2) irreducible representation basis $\vert J, J_z\rangle$. The newly introduced unitary matrix $R$ is effectively a rotation which can be defined as $R=e^{i\theta_1 J_x}e^{i\theta_2 J_y}$, rendering $\text{Tr}(\rho_{th}J_{x'})=\text{Tr}(\rho_{th}J_{y'})=0$ vanish. According to SM Section I in Ref. \cite{Murthy2023_z}, this condition indicates that $\mu_x=\mu_y=0$, and the thermal state is now as:
\begin{equation}
    U\rho_{th}U^\dagger = \frac{1}{Z} e^{-\beta(U_0HU_0^\dagger-\mu_z U_0J_z'U_0^\dagger)},\label{smeq:news}
\end{equation}
with the expected block diagonal structure.

An equivalent perspective to understand the above transformation is that the chemical potentials $\mu_a$ transforms as an adjoint representation of SU(2), namely the SU(2) symmetry of the state is understood as $G(g)\rho_{th}(\vec{\mu})G^\dagger(g) = \rho_{th}(\text{ad}(g)\vec{\mu}) $  where $g$ is the group element of SU(2) transformation and $G(g)$ is the corresponding representation operator ($R$ is an example of $G(g)$). $\text{ad}(g)$ is the adjoint representation of SU(2) ($J=1$ representation for SU(2), SO(3) faithful representation, 3D rotation). Instead, the trivial symmetry invariant definition for SU(2) symmetry $G(g)\rho_{th}G^\dagger(g) = \rho_{th}$ doesn't work for the thermal equilibrium state \cite{Marvian2014_z}, as it implies that $\text{Tr}(\rho_{th}\vec{J})=0$ which may not be compatible with the initial conditions. On the contrary, for the U(1) symmetry in the case of the commuting charges, each chemical potential is a scalar with respect to the corresponding symmetry operations, and $G(g)\rho_{th}G^\dagger(g) = \rho_{th}$ holds for $g\in U(1)$. This corresponds to the familiar definition of weak symmetry for mixed states $\rho_{th}$ which cannot directly apply to SU(2) symmetry case as we comment above. The difference lies in that SU(2) symmetry covariant state space is much larger than SU(2) symmetry invariant state space ($\langle J\rangle=0$).

Now, the density matrix after the transformation in Eq. \eqref{smeq:news} keeps the block diagonal structure across different $\{J, J_z\}$ sectors and successfully manifests the SU(2) symmetry in the system dynamics. Again, symmetry restoration is also the necessary condition for the thermalization in this non-Abelian case.

In random circuit setups, as we detailed in the following sections, the effective $H=0$ is taken for a small subsystem. For a generic random Haar circuit, the final output state in the subsystem is a fully mixed state $I/2^N$. With U(1) symmetric random circuit, the subsystem density matrix will converge to $\rho\propto e^{-\mu J_z}$ at long times limit, as long as the initial state $\rho_0$ is generic and not solely or largely lie in the charge sectors of extreme values. The equilibrium density matrix for the subsystem can be numerically checked with Eq. \eqref{eq:u1rhog}. This also explains why when the tilt angle $\theta$ is small on the ferromagnetic state, the symmetry restoration is very slow or even impossible for finite-size systems. The reason is that the initial state has a rather large overlap with $J_z=N$ sector of Hilbert space dimension $1$. Note that these $O(1)$ dimension sectors cannot thermalize as ETH doesn't apply. Therefore, the state with small $\theta$ is slow in thermalization or fails to thermalize in finite-size systems. This also explains QME: QME is observed for the set of initial states, where the less symmetry is broken, the slower thermalization and thus slower symmetry restoration set in. For the SU(2) case, since the initial state already gives zero expectation on $J_x$, $J_y$, the final thermalized subsystem follows $\rho\propto e^{-\mu J_z}$ directly. In sum, in featureless random circuit cases, due to the lack of $H$, the final thermalized subsystems are all described by the diagonal matrix. This viewpoint also underlies our assumption that $Z_2$ symmetry cannot show QME behavior as each sector of $Z_2$ symmetry is exponentially large and equally easy to thermalize. 

Specifically, for the initial states tested in SU(2) case in our work, we can directly verify that $\langle J_x\rangle=\langle J_y\rangle=0$ for even size subsystems and thus the natural direction of $J_z$ coincides with the preferred direction $J_z'$, i.e. $R=I$, only $U_0$ transformation is required to reveal the symmetry structure of the evolved state.

\section{More numerical results for U(1) symmetric quantum circuits}

\subsection{With additional symmetries}

\begin{figure}[ht]
\centering
\includegraphics[width=0.98\textwidth, keepaspectratio]{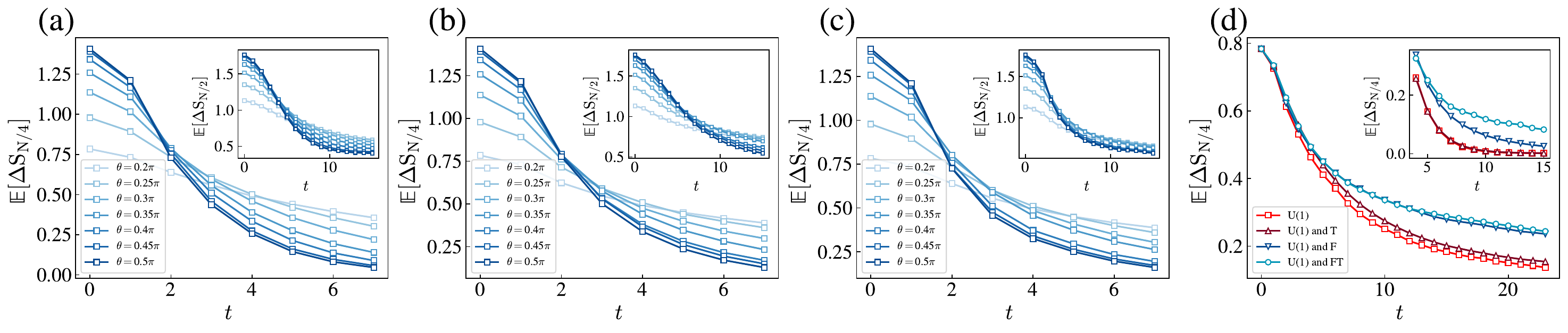}
\caption{Entanglement asymmetry dynamics with additional (a) spatial translation symmetry (T), (b) temporal translation symmetry (F), and (c) spatial and temporal translation symmetries (FT). $A$ is set to $[0, N/4]$ and $[0, N/2]$ in the main panels and the insets with $N=16$, respectively. In (d), we show the comparison between entanglement asymmetry dynamics in the presence of additional symmetries with fixed tilt angle $\theta=0.2\pi$ in the main panel and $\theta=0.5\pi$ in the inset. The presence of temporal translation symmetry slows down the U(1) symmetry restoration.}
\label{fig:EAdynamics_additionalsymmetry}
\end{figure}

In this section, we show more numerical results of the entanglement asymmetry dynamics with the same setup as that in the main text. In contrast to the previous results, the quantum circuits now possess additional symmetries beyond the U(1) symmetry. The initial state is chosen as the tilted ferromagnetic state as shown in Eq.~\textcolor{LinkColor}{2}. 
The numerical results with additional (discrete) spatial translation symmetry, temporal translation symmetry, and spatial and temporal translation symmetries are shown in Fig.~\ref{fig:EAdynamics_additionalsymmetry} (a)(b)(c) respectively. In all these cases, QME still exists. Moreover, we find that the presence of temporal translation symmetry slows down the U(1) symmetry restoration and thermalization as shown in Fig.~\ref{fig:EAdynamics_additionalsymmetry} (d). This is consistent with the results in Ref.~\cite{PhysRevB.109.024311} where U(1) symmetric random Haar Floquet circuit is reported to show slow thermalization.

\subsection{With different initial states}
\begin{figure}[ht]
\centering
\includegraphics[width=0.6\textwidth, keepaspectratio]{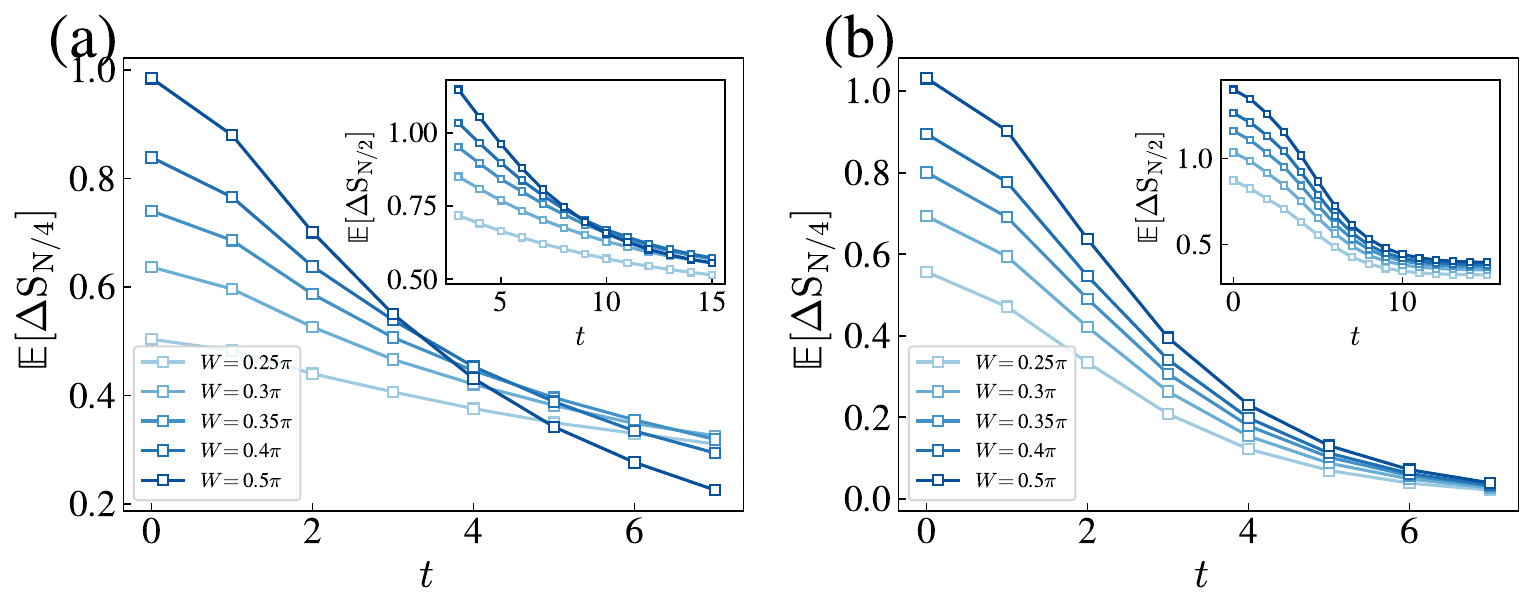}
\caption{Entanglement asymmetry dynamics with initial (a) tilted ferromagnetic state and (b) tilted N\'eel state where tilt angle on each qubit is uniformly randomly sampled from $[-W,W]$. The QME exists in the former case while it is absent in the latter case, consistent with the results of uniform tilted angles.}
\label{fig:EAdynamics_diffinit2}
\end{figure}

Besides the three sets of initial states discussed in the main text: tilted ferromagnetic states, tilted N\'eel states, and tilted ferromagnetic states with a middle domain wall, we further investigate the entanglement asymmetry dynamics with initial states chosen as tilted ferromagnetic or N\'eel state with tilt angle on each qubit randomly sampled from $[-W, W]$. Although the existence of QME depends on the choice of the initial state, QME still exists with an initial random tilted ferromagnetic state as shown in Fig.~\ref{fig:EAdynamics_diffinit2}. The presence of QME from such initial states with randomness addresses the concern that QME is only specific to some fine-tuned states. Such initial states with random parameters are more suitable for experimental demonstration of QME on quantum devices because they do not require high-precision state preparation. In other words, the on-site randomness is irrelevant to the existence of QME.

\begin{figure}[ht]
\centering
\includegraphics[width=0.6\textwidth, keepaspectratio]{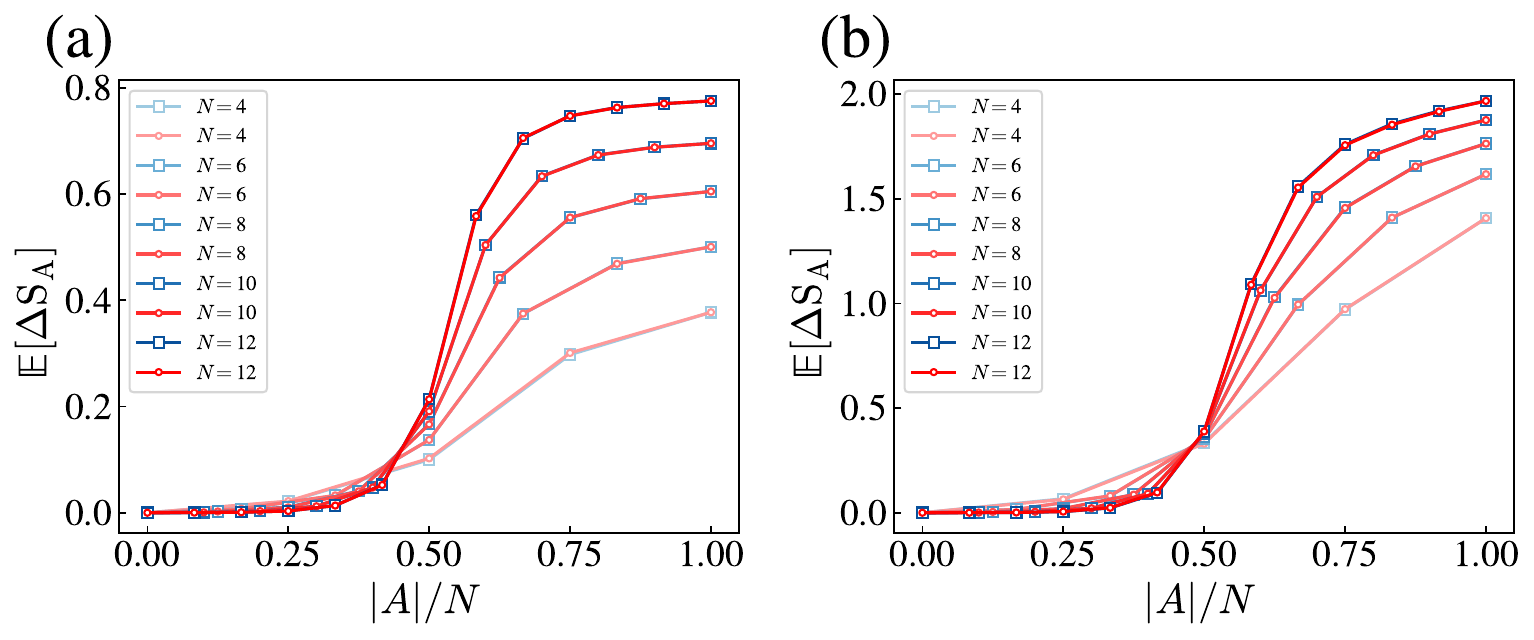}
\caption{Entanglement asymmetry at late times with initial state chosen as tilted N\'eel state (red) and tilted ferromagnetic state with a middle domain wall (blue). The tilt angles are $\theta = \pi/10$ and $\theta = \pi /2$ for (a) and (b), respectively. These two initial states have the same late-time behaviors, although the early-time behaviors are significantly different.}
\label{fig:EAlatetime_diffinit}
\end{figure}

Moreover, as mentioned in the main text, the entanglement asymmetry in the long time limit only depends on the distribution over different charge sectors of the initial states. Here, we numerically validate this prediction via the late-time entanglement asymmetry with initial states chosen as uniformly tilted N\'eel state and ferromagnetic state with a middle domain wall, the results are shown in Fig.~\ref{fig:EAlatetime_diffinit}.

\subsection{Numerical results after a quench of global U(1) symmetric unitary}
\begin{figure}[ht]
\centering
\includegraphics[width=0.85\textwidth, keepaspectratio]{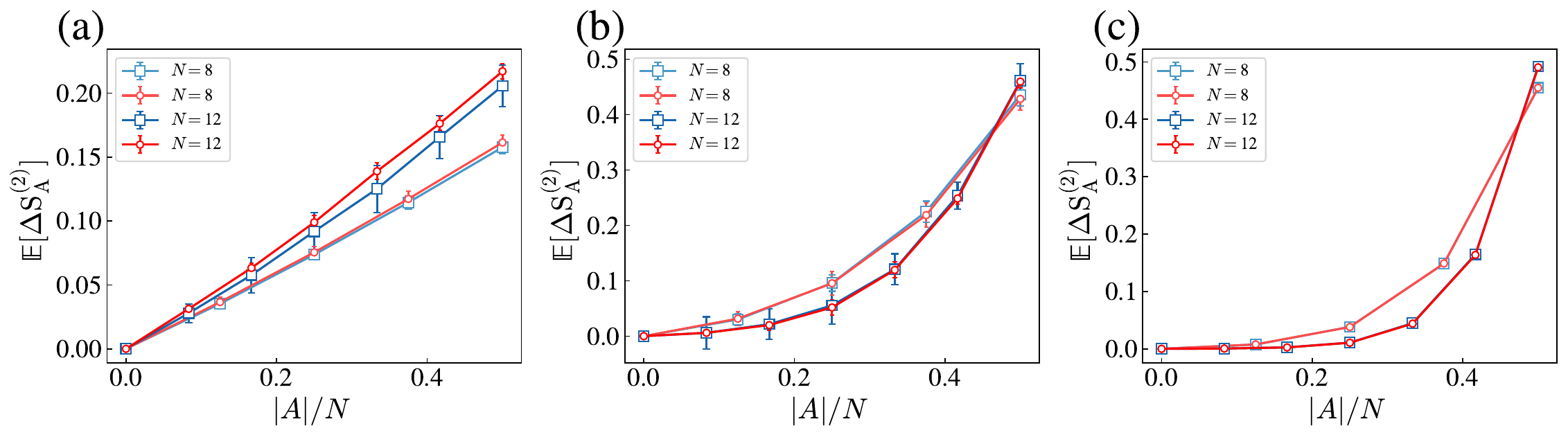}
\caption{R\'enyi-2 entanglement asymmetry at the late time of U(1) symmetric quantum circuits composed of local gates (red) and after a quench of global U(1) symmetric unitary (blue) are consistent with each other as anticipated. The initial state is chosen as a tilted ferromagnetic state with tilt angles equaling $0.1\pi$, $0.3\pi$, and $0.5\pi$ for (a)(b)(c) respectively.}
\label{fig:EA_gloabl_and_latetime}
\end{figure}

\begin{figure}[ht]
\centering
\includegraphics[width=0.6\textwidth, keepaspectratio]{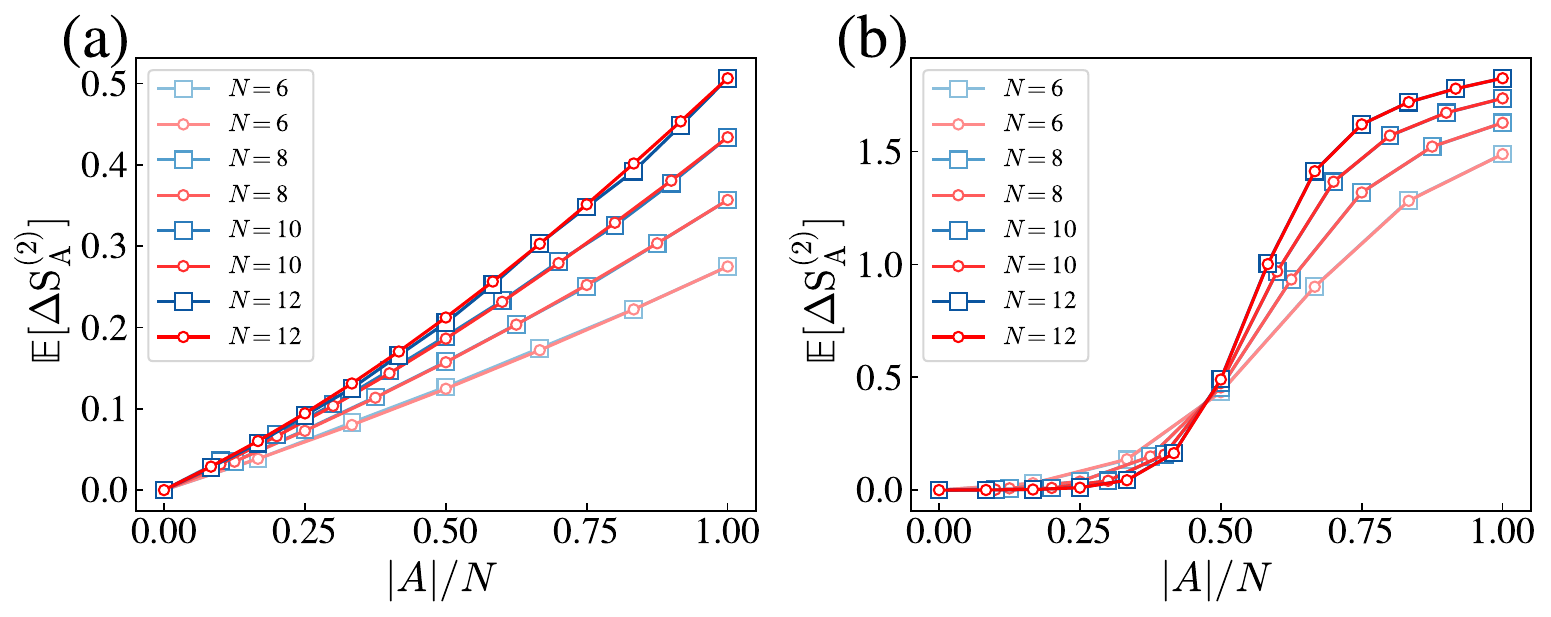}
\caption{R\'enyi-2 entanglement asymmetry after a quench of global U(1) symmetric unitary obtained from numerical simulation (blue) and analytical form (red). The initial state is chosen as a tilted ferromagnetic state with the tilt angle equaling $0.1\pi$ and $0.5\pi$ for (a) and (b) respectively.}
\label{fig:latetime_numerical_and_theory}
\end{figure}

In the analytical analysis, we have utilized a global U(1) symmetric unitary to approximate the brick-wall quantum circuits with sufficiently long depth. Here, we have numerically verified this approximation via the consistency check of R\'enyi-2 entanglement asymmetry at the late time of U(1) symmetric quantum circuits and after a quench of global U(1) symmetric gate, as shown in Fig.~\ref{fig:EA_gloabl_and_latetime}. Moreover, we have also verified the analytical results as discussed below with numerical results, as shown in Fig.~\ref{fig:latetime_numerical_and_theory}.

\subsection{Numerical results of thermalization speed}
In the main text, we have provided a unified criterion of initial states exhibiting QME: when the degree of symmetry breaking decreases, the overlap with the charge sectors of small Hilbert space dimensions also increases, thereby decelerating the rate of thermalization. In this section, we show numerical results of the thermalization speed dependence with initial states of different charge distributions to demonstrate our claim that states in the charge sector of the smaller Hilbert space dimension show slower thermalization.

\begin{figure}
\centering
\includegraphics[width=0.9\textwidth, keepaspectratio]{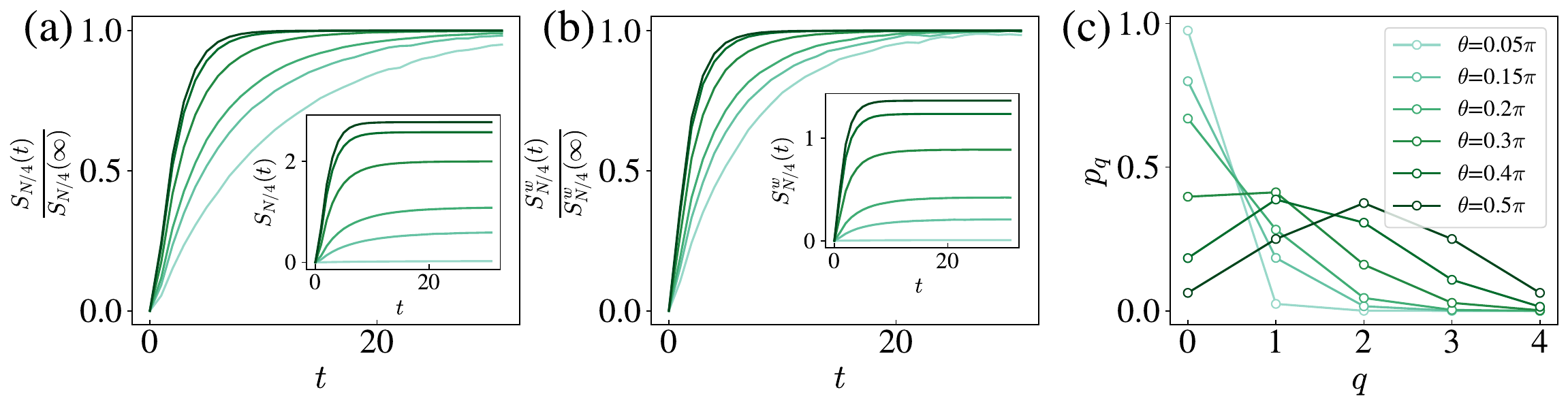}
\caption{Tilted ferromagnetic state with $N=16$ and $A=N/4$.  (a) normalized entanglement entropy dynamics (inset shows the unnormalized entanglement entropy dynamics). The increasing intensity of colors represents increasing tilt angle $\theta \in\{0.05\pi, 0.15\pi, 0.2\pi, 0.3\pi, 0.4\pi, 0.5\pi \}$. (b) normalized symmetry-resolved entanglement entropy dynamics. (c) $p_{q}$ distribution for reduced density matrix $\rho_{N/4}$ with different $\theta$.}
\label{fig:thermalization_ferro}
\end{figure}

Entanglement entropy dynamics is a metric of quantum thermalization. We calculated the von Neumann entropy $S_{A}$ of subsystem $A$ and the symmetry-resolved entanglement entropy $S^{w}_{A}$~\cite{doi:10.1126/science.aau0818, PhysRevLett.121.150501, PhysRevA.100.022324} to quantify the thermalization speed. For a given density matrix $\rho$, the reduced density matrix $\rho_{A} = {\rm tr}_{\bar{A}} \rho$ and $\rho_{A, Q} = \sum_q \Pi_q \rho_A \Pi_q = \oplus_q \rho_{A, q}$
    where $\Pi_q$ is the projector to the $q$-th charge sector of dimension $C_{\vert A \vert}^{q}$. The symmetry-resolved entanglement entropy is defined as:
    \begin{eqnarray}
        S^{w}_{A} = \sum_{q} p_{q} S^{q}_{A},
    \end{eqnarray}    
    where $S^{q}_{A} = S(\Tilde{\rho}_{A, q}) $ is the von Neumann entropy of $\Tilde{\rho}_{A, q} = \rho_{A,q} / p_{q}$ with $p_{q} =  {\rm tr} \rho_{A,q}$ is the weight of $q$-th charge sector. We note that the symmetry-resolved entanglement entropy is complementary to the entanglement asymmetry as the former measures the thermalization status within each charge sector diagonal block while the latter measures the thermalization status for the inter-charge sector off-diagonal parts. To directly compare the thermalization speeds of different sectors, we utilize the normalized $S_{A}$ and $S^{w}_{A}$ as shown in Fig.~\ref{fig:thermalization_ferro} (a) and (b). Here, the subsystem $A$ is chosen as the leftmost $N/4$ qubits under the periodic boundary conditions and other choices of $A$ do not alter the conclusions discussed below. As shown in Fig.~\ref{fig:thermalization_ferro} (c), with the decreasing of tilt angle $\theta$, i.e., the initial state is more symmetric, the charge distribution of the reduced density matrix $\rho_{A}$ is more peaked to charge sectors of small dimensions and thus the thermalization speeds decrease. Consequently, a QME occurs due to the states with lower initial entanglement asymmetry being slow to thermalize.
    
\begin{figure}
\centering
\includegraphics[width=0.9\textwidth, keepaspectratio]{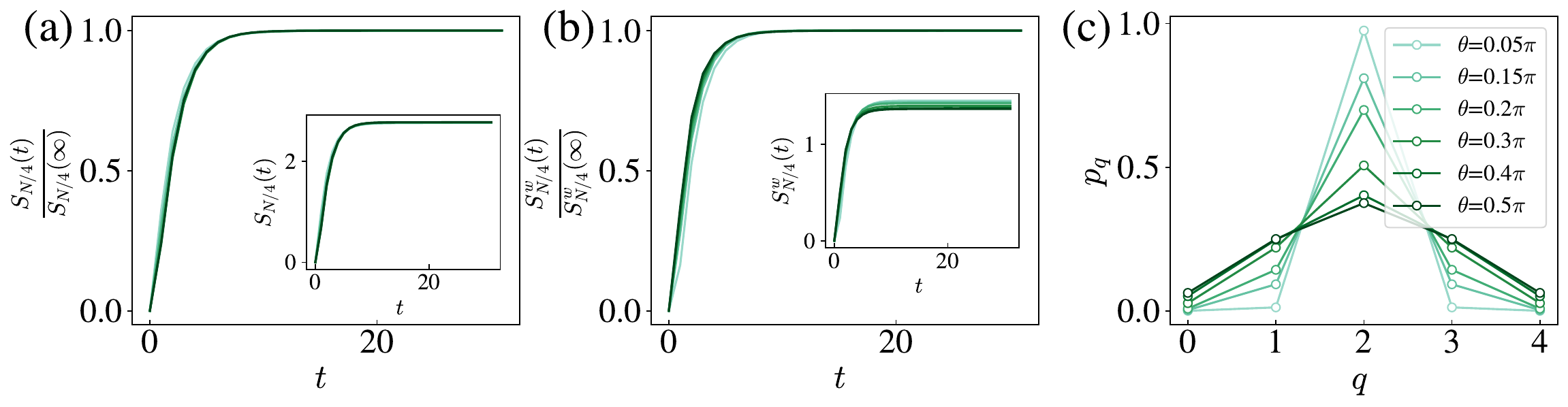}
\caption{Tilted N\'eel state with $N=16$ and $A=N/4$.  (a) normalized entanglement entropy dynamics (inset shows the unnormalized entanglement entropy dynamics). The increasing intensity of colors represents increasing tilt angle $\theta \in\{0.05\pi, 0.15\pi, 0.2\pi, 0.3\pi, 0.4\pi, 0.5\pi \}$. (b) normalized symmetry-resolved entanglement entropy dynamics. (c) $p_{q}$ distribution for reduced density matrix $\rho_{N/4}$ with different $\theta$.}
\label{fig:thermalization_neel}
\end{figure}

\begin{figure}
\centering
\includegraphics[width=0.9\textwidth, keepaspectratio]{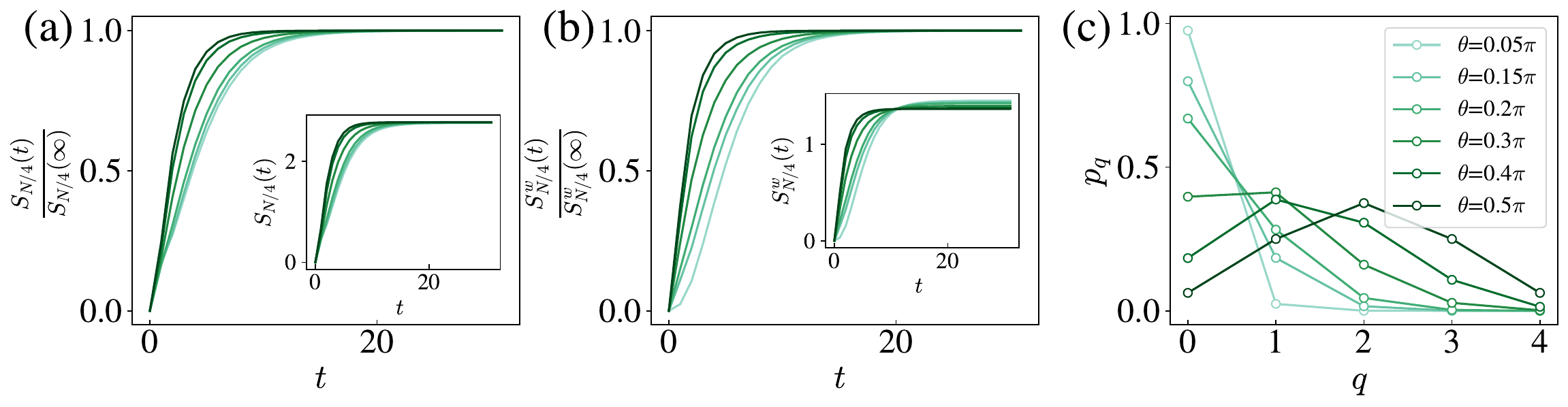}
\caption{Tilted ferromagnetic state with a middle domain wall. Here, $N=16$ and $A=N/4$.  (a) normalized entanglement entropy dynamics (inset shows the unnormalized entanglement entropy dynamics). The increasing intensity of colors represents increasing tilt angle $\theta \in\{0.05\pi,0.15\pi, 0.2\pi, 0.3\pi, 0.4\pi, 0.5\pi \}$. (b) normalized symmetry-resolved entanglement entropy dynamics. (c) $p_{q}$ distribution for reduced density matrix $\rho_{N/4}$ with different $\theta$.}
\label{fig:thermalization_half}
\end{figure}

    Besides the tilted ferromagnetic state, we have also investigated the entanglement asymmetry dynamics with initial states chosen as tilted N\'eel state
    \begin{eqnarray}
        \vert \psi_{0} (\theta)\rangle =  e^{-i\frac{\theta}{2} \sum_{j} \sigma_{j}^{y}} \vert 0101...01 \rangle,
    \end{eqnarray}
    and tilted ferromagnetic state with a middle domain wall
    \begin{eqnarray}
        \vert \psi_{0} (\theta)\rangle =  e^{-i\frac{\theta}{2} \sum_{j} \sigma_{j}^{y}} \vert 000...111 \rangle.
    \end{eqnarray}
    As shown in Fig.~\textcolor{LinkColor}{5} in the main text, the QME is absent for the former case while it is present for the latter case although the entanglement asymmetry in the long time limit should be the same. We also investigated the thermalization speeds with these different initial states to validate the theoretical prediction of QME from quantum thermalization. The numerical results for the tilted N\'eel state and tilted ferromagnetic state with a middle domain wall are shown in Fig.~\ref{fig:thermalization_neel}
    and Fig.~\ref{fig:thermalization_half} respectively. As expected, the late-time dynamics with these two different initial states are the same. However, due to the local spin configurations being different, the weights for different charge sectors differ significantly as shown in Fig.~\ref{fig:thermalization_neel} (c)
    and Fig.~\ref{fig:thermalization_half} (c). For the tilted ferromagnetic state with a middle domain wall, the $p_{q}$ distribution is similar to that of the tilted ferromagnetic state: the weights for charge sectors with smaller Hilbert dimensions increase with the tilted angle $\theta$ decrease. Consequently, the more symmetric initial state has a slower thermalization speed and thus the QME occurs. On the contrary, for the tilted N\'eel state, the charge distribution is more peaked to the charge sector of the largest Hilbert space dimension as the tilted angle $\theta$ decreases. Therefore, the tilted N\'eel state strongly obeys the ETH regardless of the tilted angle $\theta$ and thus the speed of thermalization is essentially unchanged. Consequently, the entanglement asymmetry remains larger for a more asymmetric initial state and the QME is absent.

\begin{figure}
\centering
\includegraphics[width=0.9\textwidth, keepaspectratio]{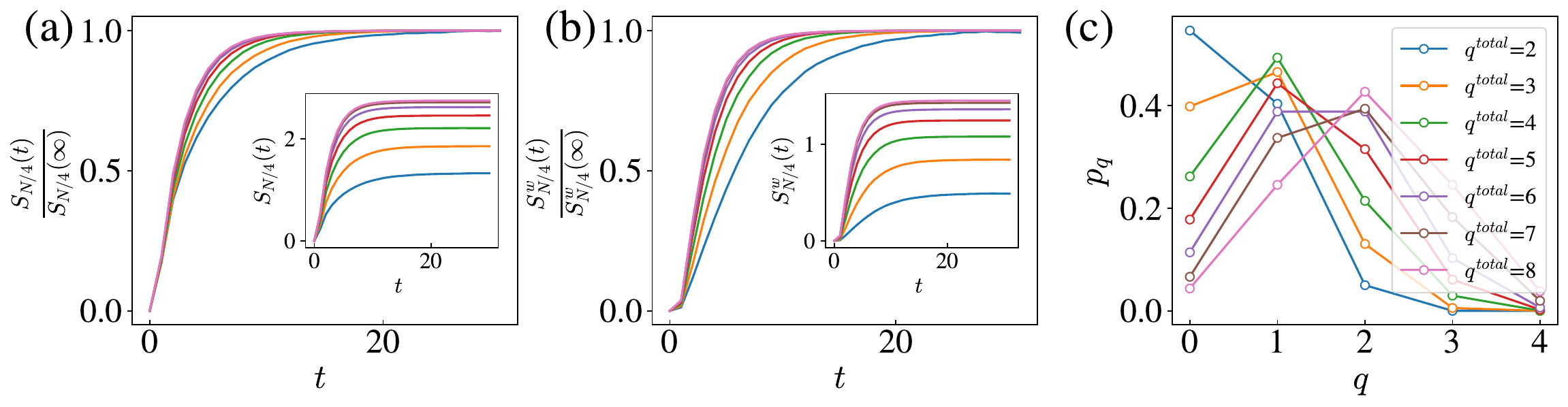}
\caption{Random bitsring with fixed total charge $q^{total}$ as initial states. Here, $N=16$ and $A=N/4$.  (a) normalized entanglement entropy dynamics (inset shows the unnormalized entanglement entropy dynamics). (b) normalized symmetry-resolved entanglement entropy dynamics. (c) $p_{q}$ distribution for reduced density matrix $\rho_{N/4}$ with different total charge $q^{total}$. We have excluded charge sectors $q^{total}=0,1$ because the symmetry-resolved entanglement entropy is zero.}
\label{fig:random_bitstring}
\end{figure}

    To further validate the conjecture that the thermalization speed is slower for the charge sector with a smaller Hilbert space dimension, we choose random bitstring with fixed charge $q^{total}$ as initial states. The entanglement entropy and symmetry-resolved entanglement entropy dynamics shown in Fig.~\ref{fig:random_bitstring} demonstrate the claim above. Based on these convincing numerical results, we have fully established the connection between the effective Hilbert space dimension, the QME, and the thermalization effects. More importantly, this connection is general, regardless of the choice of chaotic symmetric Hamiltonian.

\subsection{Universal feature of symmetry-breaking behavior with small tilt angle $\theta$}
In the main text, we have mentioned the symmetry-breaking behavior at late times with a small tilt angle $\theta$ under the evolution of U(1)-symmetric random circuits. We comment that the feature is universal for generic chaotic evolution in finite-size systems. Theoretically speaking, when the mean charge of the initial state $\sim 1/\sqrt{N}$ is comparable to its fluctuations, the local density matrix will never agree with the grand-canonical result, since the binomial distribution of charges is not well approximated by a Gaussian distribution. We have performed additional numerical simulations with general U(1)-symmetric interacting Hamiltonians to demonstrate the symmetry breaking behavior with small tilt angle $\theta$ is a general feature
of U(1)-restoring dynamics.

\begin{figure}
\centering
\includegraphics[width=0.8\textwidth, keepaspectratio]{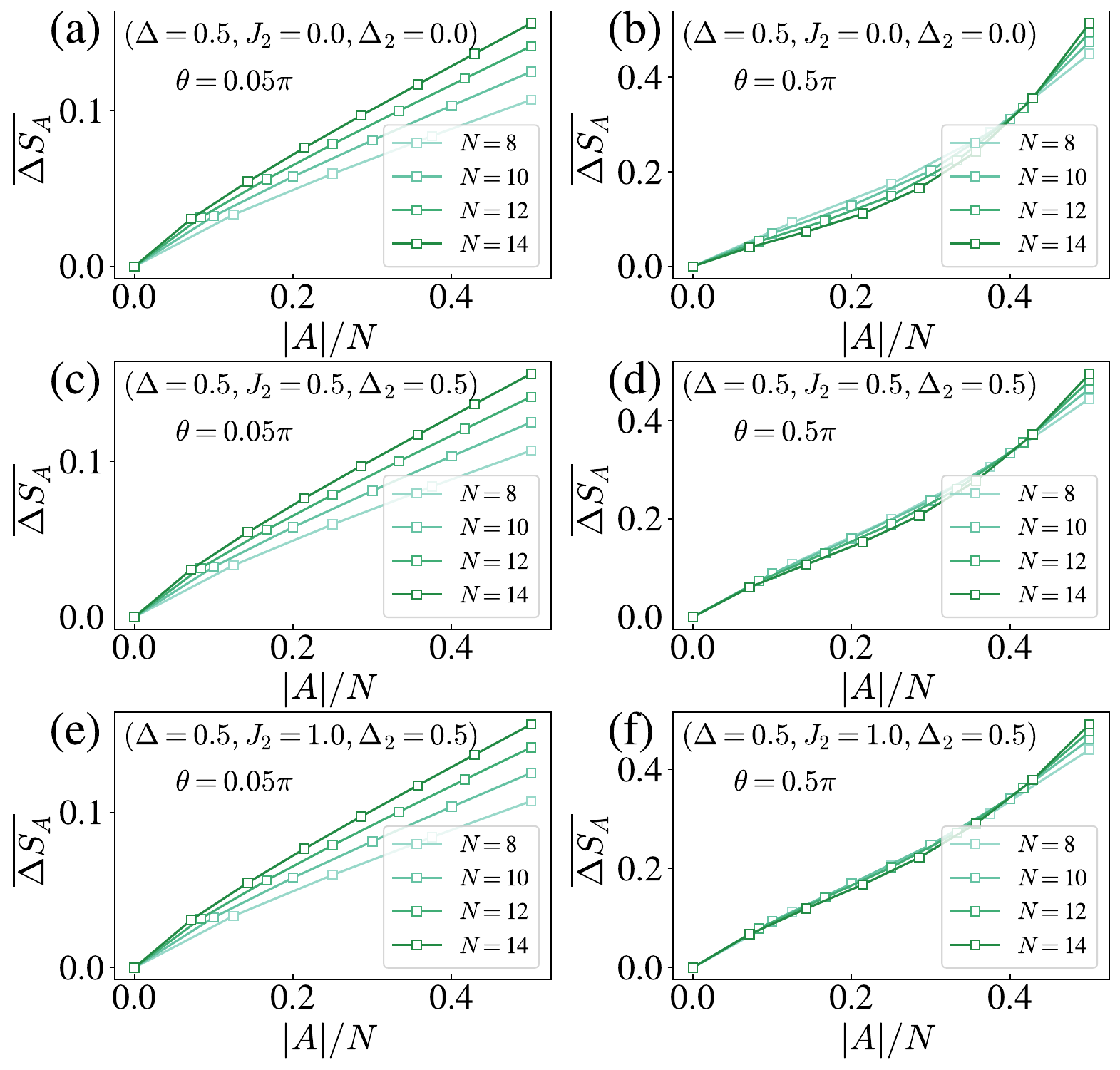}
\caption{The entanglement asymmetry at late times for different Hamiltonians and different $\theta$. Left: $\theta=0.05 \pi$; right: $\theta=0.5 \pi$. The symmetry-breaking behaviors always exist when $\theta$ is small. When $\theta$ is large, although EA approaches zero as the system size increases, we note that the finite-size effect is severe with the introduction of non-integrability.}
\label{fig:difftheta_diffH}
\end{figure}

\begin{figure}
\centering
\includegraphics[width=0.8\textwidth, keepaspectratio]{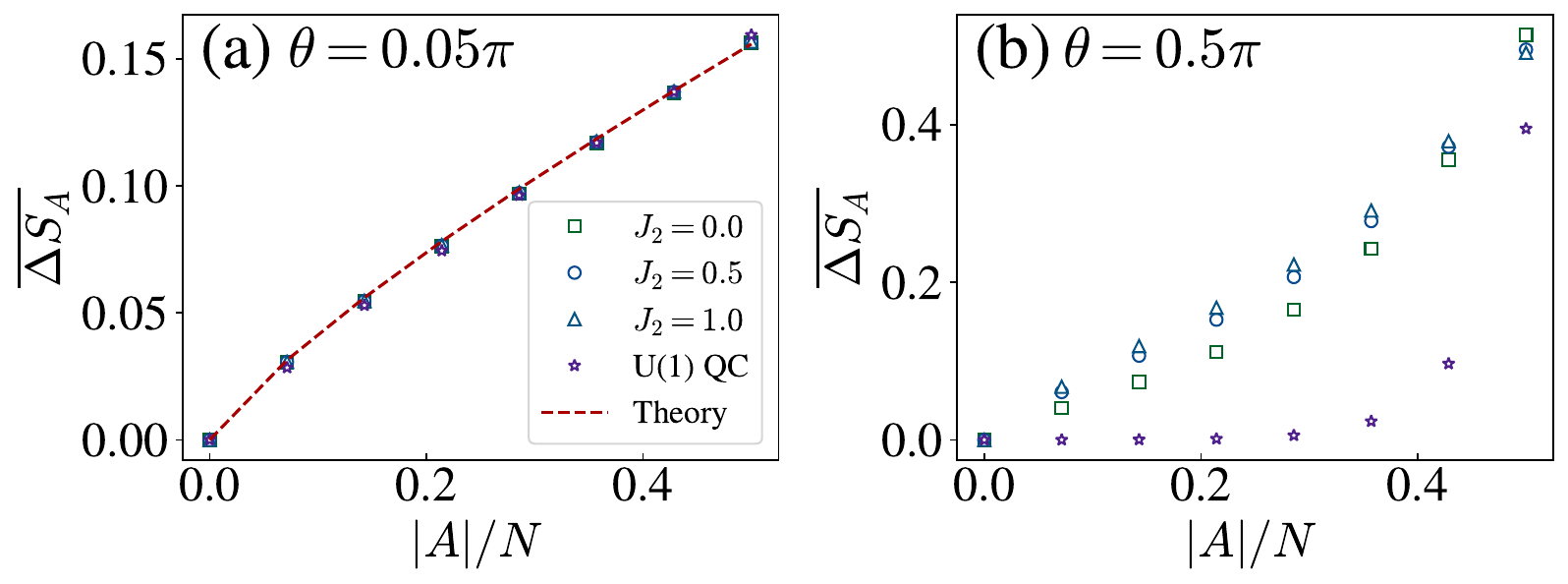}
\caption{The comparison between the late-time EA under the quench of different Hamiltonians and U(1)-symmetric quantum circuits. The Hamiltonian parameters are set as those in Fig.~\ref{fig:difftheta_diffH} and system size $N=14$. When $\theta$ is small, EA doesn't depend on the choice of Hamiltonian or quantum circuits (other numerical results with smaller system sizes supporting this conclusion are not shown here) and corresponds to that of $\vert \psi^{\prime} \rangle$.}
\label{fig:J20_J20.5_J21.0_u1qc}
\end{figure}

The testbed Hamiltonian is chosen as 
\begin{eqnarray}
    H = -\frac{1}{4} \sum_{j=1}^{N} \left[ \sigma^{x}_{j} \sigma^{x}_{j+1} + \sigma^{y}_{j} \sigma^{y}_{j+1} + \Delta \sigma^{z}_{j} \sigma^{z}_{j+1}  \right] -\frac{J_2}{4} \sum_{j=1}^{N} \left[ \sigma^{x}_{j} \sigma^{x}_{j+2} + \sigma^{y}_{j} \sigma^{y}_{j+2} + \Delta_{2} \sigma^{z}_{j} \sigma^{z}_{j+2}  \right],
    \label{eq:H}
\end{eqnarray} 
with periodic boundary conditions. The model is integrable when $J_{2}=0$ and the introduction of nonzero $J_{2}$ will break the integrability, leading to a generic quantum chaotic Hamiltonian. The initial state is chosen as the tilted ferromagnetic state utilized in the main text. Under the quench of this Hamiltonian, the entanglement asymmetry is not monotonically decreasing over time and will oscillate at late times in finite-size systems as discussed in Ref.~\cite{aresEntanglementAsymmetryProbe2023}. Therefore, we use the average of the entanglement asymmetry at late times to quantify the symmetry restoration in finite-size systems, which is given by
\begin{eqnarray}
     \overline{\Delta S_{A}} &=& \frac{1}{t_{2} - t_{1}} \int_{t_{1}}^{t_{2}} \Delta S_{A} (t) dt \\ \nonumber
    &\approx& \frac{1}{t_{2} - t_{1}} \sum_{t=t_{1}}^{t_{2}} \Delta S_{A}(t) \Delta t.
\end{eqnarray}
Here, we set $\Delta t =1$, $t_{1}=10^3$, and $t_{2}=N \times 10^4$. When tilt angle $\theta$ is sufficiently small, as illustrated in Fig.~\ref{fig:difftheta_diffH}, the symmetry-breaking behaviors at late times always exist (larger size system has larger averaged entanglement asymmetry at late times), regardless of the choices of Hamiltonian parameters. On the contrary, when $\theta$ is large, entanglement asymmetry approaches zero as the system size increases. Moreover, as shown in Fig.~\ref{fig:J20_J20.5_J21.0_u1qc} (b), we note that the entanglement asymmetry dynamics in random circuits have a much smaller finite-size effect than Hamiltonian evolution. This fact further justifies the effectiveness of the random circuits testbed utilized in our work. 

When $\theta$ is sufficiently small, the tilted ferromagnetic state with finite size can be approximated as:
\begin{eqnarray}
    \label{eq:approximation}
    \vert \psi_{0} \rangle &=& (\cos \frac{\theta}{2} \vert 0 \rangle + \sin \frac{\theta}{2} \vert 1 \rangle)^{\otimes N}  \\ \nonumber
    &\approx& \frac{1}{Z} \left[ (\cos \frac{\theta}{2} )^N \vert 0 \rangle^{\otimes N} + (\cos \frac{\theta}{2} )^{N-1} \sin \frac{\theta}{2} \sum_{i} X_{i} \vert 0 \rangle^{\otimes N}  \right] \\ \nonumber
    &=& \vert \psi^{\prime} \rangle \\ \nonumber
    &=& c_{0} \vert \psi^{q=0} \rangle +  c_{1} \vert \psi^{q=1} \rangle,
\end{eqnarray}
where $Z$ is the normalization factor. $\vert \psi^{\prime} \rangle $ is a superposition of $\vert \psi^{q=0} \rangle$ and $\vert \psi^{q=1} \rangle$, living in $q=0$ and $q=1$ charge sectors respectively. It is easy to show that $\vert \psi^{q=0} \rangle$ and  $\vert \psi^{q =1} \rangle$ are both eigenstates of the Hamiltonian in Eq.~\eqref{eq:H}. Consequently,
\begin{eqnarray}
    \vert \psi^{\prime}(t) \rangle = e^{-iHt} \vert \psi^{\prime} \rangle = c_{0} \vert \psi^{q=0} \rangle +  c_{1}e^{-i\varepsilon t} \vert \psi^{q=1} \rangle,
\end{eqnarray}
up to a global phase factor, where $\varepsilon$ is the eigenenergy of $\vert \psi^{q=1} \rangle$. This relative phase does not affect the entanglement asymmetry. Therefore, entanglement asymmetry remains the same as that of the initial state, regardless of the details of the U(1)-symmetric quench, as illustrated in Fig.~\ref{fig:J20_J20.5_J21.0_u1qc}. For the case of U(1)-symmetric random circuits, we find the entanglement asymmetry at late times is also almost the same as that obtained from $\vert \psi^{\prime} \rangle$ as shown in Fig.~\ref{fig:J20_J20.5_J21.0_u1qc}. Obviously, the approximation shown in Eq.~\ref{eq:approximation} fails in the thermodynamic limit and thus this symmetry-breaking behavior is a finite-size effect. In conclusion, the persistent symmetry breaking in finite size systems is a general feature of U(1)-restoring dynamics quenching staring from a weak symmetry-breaking tilted ferromagnetic state.

Because the thermodynamic limit has been directly taken in the derivation of the analytical results of the entanglement asymmetry dynamics under the quench of XX Hamiltonian~\cite{aresEntanglementAsymmetryProbe2023}, persistent symmetry breaking, the general and experimental relevant feature, has not been identified and investigated before. The findings on the success and failure of symmetry restoration presented in this work extend beyond theoretical insights into random circuit dynamics and offer valuable perspectives for understanding symmetry restoration in the context of more general symmetric Hamiltonians. 

\section{Numerical results for other symmetric quantum circuits}
\subsection{$Z_{2}$ symmetric quantum circuits}
\begin{figure}[ht]
\centering
\includegraphics[width=0.7\textwidth, keepaspectratio]{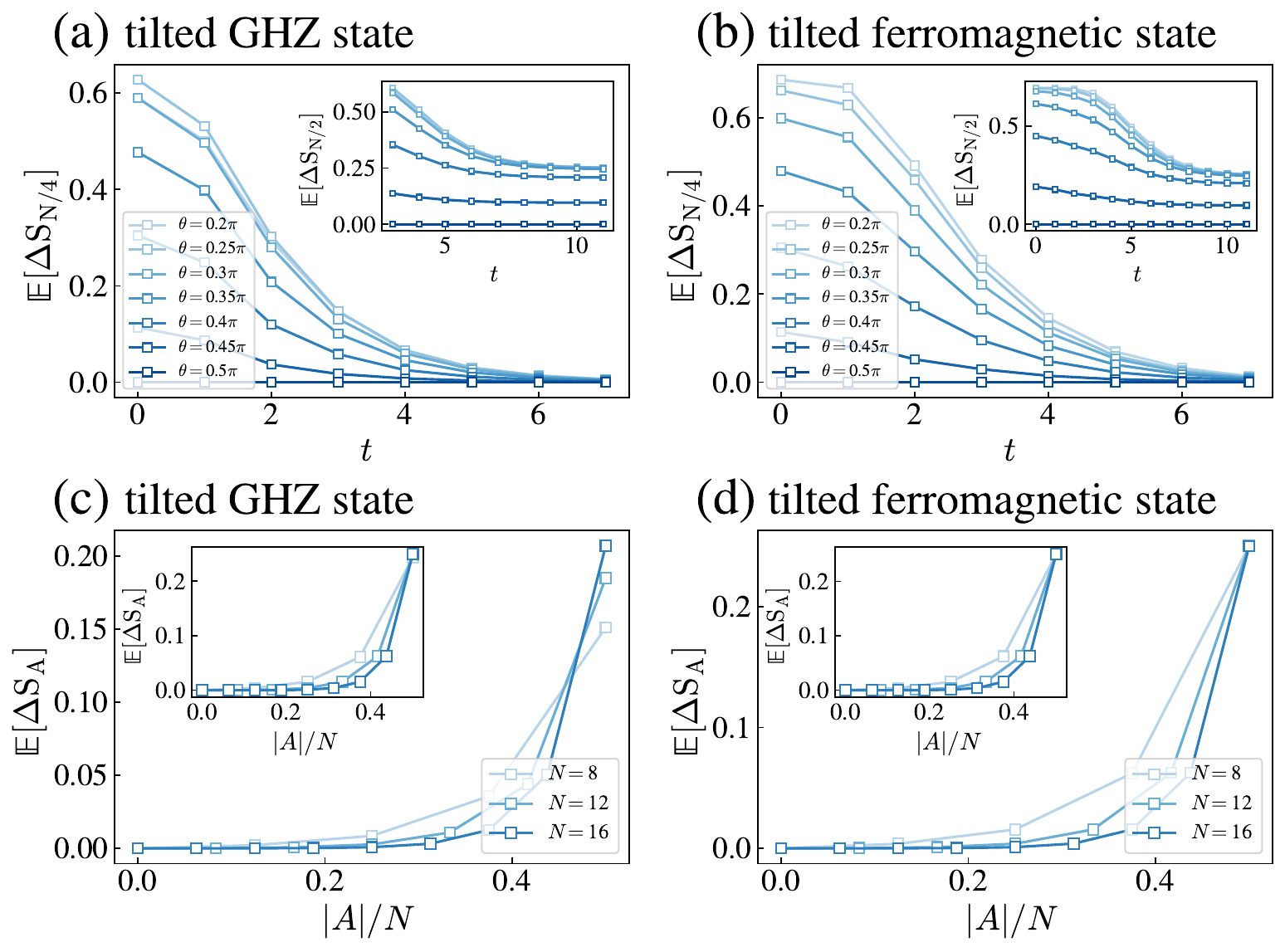}
\caption{(a)(b) show the entanglement asymmetry dynamics of $Z_{2}$ symmetric circuits with $N=16$. (c)(d) show the entanglement asymmetry at late times versus subsystem size. The tilt angles are $\theta=0.1 \pi$ and $\theta=0.2 \pi$ in the main panel and the inset. The initial state is the tilted GHZ state and the tilted ferromagnetic state for (a)(c) and (b)(d) respectively.}
\label{fig:EA_Z2}
\end{figure}

In this section, we present the numerical results of entanglement asymmetry dynamics in the $Z_{2}$ symmetric quantum circuits. The symmetry operator of $Z_{2}$ symmetry is $\prod_{i}\sigma_{i}^{x}$ and $Z_{2}$ symmetric two-qubit gates are given by 
\begin{eqnarray}
    U_{Z_2} = T^{\dag} \begin{pmatrix} U_{+1} & 0 \\ 0 & U_{-1} \end{pmatrix} T,
\end{eqnarray}
where $U_{+1}$ and $U_{-1}$ are $2 \times 2$ random Haar matrix corresponding to the $+1$ and $-1$ eigenbasis respectively, and $T$ is the transformation matrix from $Z_{2}$ symmetric basis to $z$ basis,
\begin{eqnarray}
    T = \begin{pmatrix} \frac{1}{\sqrt{2}} & 0 & 0 & \frac{1}{\sqrt{2}} \\
    0 & \frac{1}{\sqrt{2}} & \frac{1}{\sqrt{2}} & 0 \\
    0 & \frac{1}{\sqrt{2}} & \frac{-1}{\sqrt{2}} & 0 \\
    \frac{1}{\sqrt{2}} & 0 & 0 & \frac{-1}{\sqrt{2}}
    \end{pmatrix}.
\end{eqnarray}
We choose two different initial states: one is the tilted GHZ state given by
\begin{eqnarray}
    \vert \psi_{0}(\theta) \rangle = e^{-i \frac{\theta}{2} \sum_{j} \sigma^{y}_{j}} \vert \rm{GHZ} \rangle,
\end{eqnarray}
where $\vert \rm{GHZ} \rangle = \frac{1}{\sqrt{2}} (\vert 0 \rangle^{\bigotimes N} + \vert 1 \rangle^{\bigotimes N})$. The other is the tilted ferromagnetic state. As shown in Fig.~\ref{fig:EA_Z2} (a)(b), there is no QME in the entanglement asymmetry dynamics of $Z_{2}$ symmetric quantum circuits. Moreover, we also calculate the entanglement asymmetry at late times. Different from the case of U(1) symmetric circuits with a tilted ferromagnetic state as the initial state, the $Z_{2}$ symmetry of subsystem $A$ with $\vert A \vert < N/2$ will always be restored, independent of the choice of initial states and tilt angles, as shown in Fig.~\ref{fig:EA_Z2} (c)(d).

\subsection{SU(2) symmetric quantum circuits}
\begin{figure}[ht]
\centering
\includegraphics[width=0.45\textwidth, keepaspectratio]{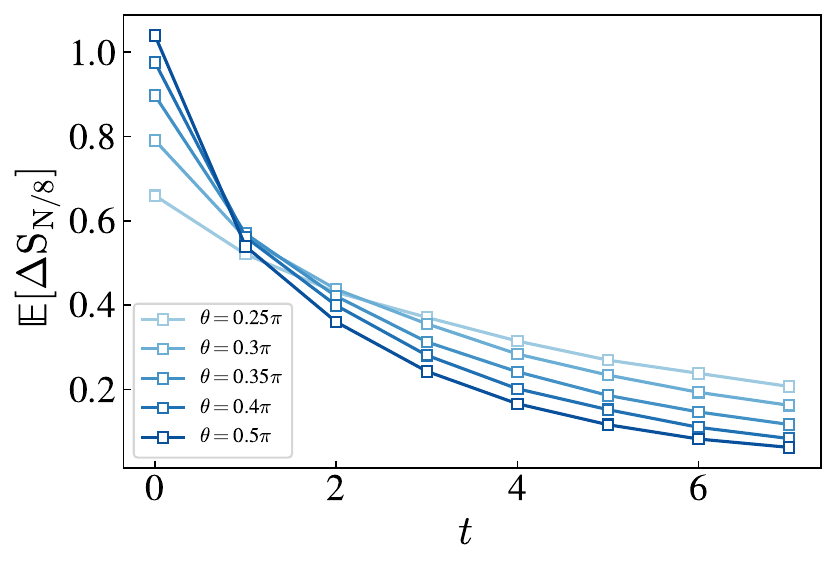}
\caption{Entanglement asymmetry dynamics of SU(2) symmetric quantum circuits with the presence of QME. Here $N=16$. The initial state is a tilted ferromagnetic state with staggered tilt angles.}
\label{fig:EAdynamics_SU2}
\end{figure}

In the main text, we have demonstrated that the QME emerges in the entanglement asymmetry dynamics of U(1) symmetric quantum circuits. Here, we show numerical results for the SU(2) symmetric quantum circuits. The initial state is chosen as the tilted ferromagnetic state with staggered tilt angles, i.e.,
\begin{eqnarray}
    \vert \psi_{0}(\theta) \rangle = e^{-i\frac{\theta}{2} \sum_{j} (-1)^{j} \sigma_{j}^{y} } \vert 000...0 \rangle.
\end{eqnarray}
We can directly verify that $\langle J_x \rangle = \langle J_y \rangle = 0$ for even size subsystems and thus the natural direction of $z$ coincides with the preferred direction $z^{\prime}$ for the block diagonalization of the equilibrium density matrix. Therefore, the SU(2) symmetry restoration indicates that the reduced density matrix is block-diagonal corresponding to the $\{ J^{2}, J_{z} \}$ sectors. However, we emphasize that the block-diagonal structure is only a necessary but not sufficient condition for the non-Abelian symmetric density matrix because the Schur lemma also requires there is some equivalence among different sectors when the dimension of the irreducible representation is larger than $1$. Namely, different from the case of Abelian symmetries, the zero entanglement asymmetry does not necessarily mean that the quantum state respects the corresponding non-Abelian symmetry. In sum, vanishing entanglement asymmetry with repect to SU(2) symmetry is a necessary condition for SU(2) symmetry restoration and SU(2) symmetry restoration is a necessary condition for quantum thermalization. Therefore, we can still use the entanglement asymmetry as an indicator of the symmetry restoration and thermalization for the non-Abelian symmetries.
The numerical results are shown in Fig.~\ref{fig:EAdynamics_SU2}. We also observe the QME in the entanglement asymmetry dynamics as anticipated by the unified mechanism we developed in the main text.

\section{Analytical Results}

In this section, we provide the analytical results for the dynamical and steady behaviors of entanglement asymmetry under the evolution of random quantum circuits with and without U(1) symmetry.

\subsection{Calculation without symmetry in the early time}
As shown in the main text, the entanglement asymmetry dynamics collapse for different initial states for random circuits without any symmetry. To understand this phenomenon, we can consider the average of random two-qubit gates,
\begin{eqnarray}
    \mathbb{E}_{\mathcal{U}}(U_{t,(i,i+1)} \otimes U_{t,(i,i+1)}^{*})^{\otimes 2} = \sum_{\sigma, \tau \in S_{2}} \text{Wg}_{d^2}^{(2)}(\sigma \tau^{-1}) \vert \tau \tau \rangle \langle \sigma \sigma \vert_{i,i+1}, 
\end{eqnarray}
where $S_{2} = \{I,S \}$ with $I=\sum_{ij} \vert ii jj \rangle$ and $S=\sum_{ij} \vert ijji \rangle$ being the elements from $S_2$ permutation group. The initial ferromagnetic state can be written as $\vert \psi_{0} (\theta) \rangle = \bigotimes_{i} \psi_{0}^{i} (\theta)$ where $ \vert \psi_{0}^{i} (\theta) \rangle = \cos \frac{\theta}{2} \vert 0 \rangle_{i} + \sin \frac{\theta}{2} \vert 1 \rangle_{i} $. The reduced density matrix in the two-copy replicated Hilbert space is 
\begin{eqnarray}
\vert \rho_{i} \rangle = (\cos^{2} \frac{\theta}{2} \vert 00 \rangle + \cos \frac{\theta}{2} \sin \frac{\theta}{2} \vert 01 \rangle + \cos \frac{\theta}{2} \sin \frac{\theta}{2} \vert 10 \rangle + \sin^{2} \frac{\theta}{2} \vert 11 \rangle )^{\otimes 2}.   
\end{eqnarray}
Thus, 
\begin{eqnarray}
    \langle I \vert \rho_{i} \rangle = \langle S \vert \rho_{i} \rangle = 1.
\end{eqnarray}

Therefore, the output state of the first layer of random two-qubit gates does not depend on the initial state, and the entanglement asymmetry dynamics collapse for different initial states. In fact, this conclusion still holds as long as the initial state is a product state.
\newline ~ \newline

Next, we provide the analytical results for the dynamics of entanglement asymmetry under the evolution of random quantum circuits with and without U(1) symmetry in the long-time limit. 

\subsection{Calculation without symmetry in the long-time limit}

Suppose the evolution time is sufficiently long, e.g. polynomial with the system size. In that case, the circuit ensemble can usually be approximated by a single Haar-random unitary acting on all qubits at least for the second moment in R\'enyi-2 entanglement asymmetry~\cite{Brandao2016_k}. We denote this ``global'' random unitary as $U$ and the corresponding ensemble as $\mathbb{U}$. Suppose $\rho_0$ is a pure initial state and $\rho=U^\dagger \rho_0 U$ is the resulting randomly evolved state. Denote the reduced density matrix of subsystem $A$ as $\rho_{A} = {\rm tr}_{\bar{A}}(\rho)$, where $\bar{A}$ is the complement region of $A$. We use $|A|$ to represent the number of qubits in the subsystem $A$. The total number of qubits is denoted as $N$.

If there is no symmetry restriction at all, then $\mathbb{U}$ just means the Haar measure on the whole Hilbert space. The corresponding detailed calculation can be found in Ref.~\cite{ares2023entanglement}. 
Here we reproduce the same results using an alternative method, which facilitates the derivation of the U(1)-symmetric case below. As we know, the purity of $\rho_A$ can be written as the expectation of a certain SWAP operator with respect to the 2-fold replica of $\rho$, i.e.,
\begin{equation}
    {\rm tr}(\rho_A^2) = {\rm tr}_{2A}(\rho_A^{\otimes 2} S\vert_{2A}) = {\rm tr}_{2A}[{\rm tr}_{2\bar{A}}(\rho^{\otimes 2}) S\vert_{2A}] = {\rm tr}[\rho^{\otimes 2} (S\vert_{2A}\otimes I\vert_{2\bar{A}})],
\end{equation}
where $S\vert_{2A}$ and $I\vert_{2\bar{A}}$ are the SWAP operator and the identity on the 2-fold replicas of subsystem $A$ and $\bar{A}$, respectively. The corresponding tensor network diagram is
\begin{equation}
\begin{mytikz4}
\draw    (270,85.03) -- (270,166.37) ;
\draw    (230.4,110.28) -- (230.4,166.03) ;
\draw  [fill={rgb, 255:red, 204; green, 230; blue, 255 }  ,fill opacity=1 ] (280,126.6) .. controls (280,126.6) and (280,126.6) .. (280,126.6) -- (280,146.6) .. controls (280,146.6) and (280,146.6) .. (280,146.6) -- (220,146.6) .. controls (220,146.6) and (220,146.6) .. (220,146.6) -- (220,126.6) .. controls (220,126.6) and (220,126.6) .. (220,126.6) -- cycle ;
\draw    (254.67,94.03) -- (254.67,173.03) ;
\draw    (215.07,116.95) -- (215.07,172.7) ;
\draw  [fill={rgb, 255:red, 204; green, 230; blue, 255 }  ,fill opacity=1 ] (264.67,133.27) .. controls (264.67,133.27) and (264.67,133.27) .. (264.67,133.27) -- (264.67,153.27) .. controls (264.67,153.27) and (264.67,153.27) .. (264.67,153.27) -- (204.67,153.27) .. controls (204.67,153.27) and (204.67,153.27) .. (204.67,153.27) -- (204.67,133.27) .. controls (204.67,133.27) and (204.67,133.27) .. (204.67,133.27) -- cycle ;
\draw    (215.07,116.95) .. controls (215.2,103.7) and (229.87,99.7) .. (230.2,84.37) ;
\draw    (230.4,110.28) .. controls (230.53,97.03) and (215.2,107.7) .. (215.53,92.37) ;
\draw [color={rgb, 255:red, 0; green, 0; blue, 0 }  ,draw opacity=0.27 ] [dash pattern={on 3pt off 1.5pt}]  (215.53,89.48) .. controls (212.87,77.73) and (180.53,74.43) .. (180.87,132.77) .. controls (181.2,191.1) and (211.87,191.84) .. (215.07,177.9) ;
\draw [color={rgb, 255:red, 0; green, 0; blue, 0 }  ,draw opacity=0.27 ] [dash pattern={on 3pt off 1.5pt}]  (230.2,80.67) .. controls (227.53,68.93) and (195.2,65.63) .. (195.53,123.96) .. controls (195.87,182.3) and (227.2,184.5) .. (230.4,170.56) ;
\draw [color={rgb, 255:red, 0; green, 0; blue, 0 }  ,draw opacity=0.27 ] [dash pattern={on 3pt off 1.5pt}]  (254.45,89.81) .. controls (257.32,78.07) and (292.22,74.77) .. (291.86,133.1) .. controls (291.5,191.44) and (258.4,192.17) .. (254.95,178.23) ;
\draw [color={rgb, 255:red, 0; green, 0; blue, 0 }  ,draw opacity=0.27 ] [dash pattern={on 3pt off 1.5pt}]  (270.28,80.34) .. controls (273.16,68.6) and (308.06,65.29) .. (307.7,123.63) .. controls (307.34,181.97) and (273.52,184.17) .. (270.07,170.23) ;

\end{mytikz4}\quad.
\end{equation}
Using the formula from the Weingarten calculus~\cite{Collins2006_k}, the average of the 2-fold replica of the Haar-random state is
\begin{equation}
    \mathbb{E}_\mathbb{U} [\rho^{\otimes 2}] = \mathbb{E}_\mathbb{U} [ U^{\dg \otimes 2} \rho_0^{\otimes 2} U^{\otimes 2}] = \frac{ ({\rm tr}\rho_0)^2 I + {\rm tr}(\rho_0^2) S}{d^2-1} - \frac{ ({\rm tr}\rho_0)^2 S + {\rm tr}(\rho_0^2) I}{d(d^2-1)} =  \frac{I + S}{d(d+1)},
\end{equation}
where we have used the fact that $\rho_0$ is a pure state. Here $d=2^N$ denotes the Hilbert space dimension of the entire system. Hence, the average of the purity of $\rho_A$ is
\begin{equation}
    \mathbb{E}_\mathbb{U} [{\rm tr}(\rho_A^2)] = {\rm tr}[\frac{(I+S)}{d(d+1)}(S\vert_{2A}\otimes I\vert_{2\bar{A}})] = \frac{d_A d_{\bar{A}}^2 + d_{\bar{A}} d_A^2}{d(d+1)} = \frac{ d_{\bar{A}} + d_A}{d+1} = \frac{2^{|\bar{A}|}+2^{|A|}}{2^N+1},
\end{equation}
where $d_A$ and $d_{\bar{A}}$ denote the Hilbert space dimensions of subsystem $A$ and $\bar{A}$ respectively. Next, we consider the ``pruned'' reduced state~\cite{aresEntanglementAsymmetryProbe2023}
\begin{equation}
    \rho_{A,Q} = \sum_{q=0}^{|A|} \Pi_q\vert_{A} \rho_A \Pi_q\vert_{A},
\end{equation}
where $\Pi_q\vert_{A}$ is the projector on the charge sector of $q$ charges restricted to subsystem $A$. Namely, $\rho_{A,Q}$ is the result of removing all non-zero elements outside the block-diagonal structure of charge sectors in $\rho_A$. The purity of $\rho_{A,Q}$ can be written as
\begin{equation}
\begin{aligned}
    {\rm tr}(\rho_{A,Q}^2) &= \sum_{q=0}^{|A|} {\rm tr}\left[\Pi_q\vert_{A} \rho_A \Pi_q\vert_{A} \rho_A \Pi_q\vert_{A} \right] = \sum_{q=0}^{|A|} {\rm tr}\left[ \rho^{\otimes 2} ((\Pi_q\vert_{A})^{\otimes 2} S\vert_{2A}) \otimes I\vert_{2\bar{A}} \right] .
\end{aligned}
\end{equation}
The tensor network diagram corresponding to a single term in the summation is
\begin{equation}
\begin{mytikz4}
\draw    (270,85.03) -- (270,166.37) ;
\draw    (230.4,110.28) -- (230.4,166.03) ;
\draw  [fill={rgb, 255:red, 204; green, 230; blue, 255 }  ,fill opacity=1 ] (280,126.6) .. controls (280,126.6) and (280,126.6) .. (280,126.6) -- (280,146.6) .. controls (280,146.6) and (280,146.6) .. (280,146.6) -- (220,146.6) .. controls (220,146.6) and (220,146.6) .. (220,146.6) -- (220,126.6) .. controls (220,126.6) and (220,126.6) .. (220,126.6) -- cycle ;
\draw    (254.67,94.03) -- (254.67,173.03) ;
\draw    (215.07,116.95) -- (215.07,172.7) ;
\draw  [fill={rgb, 255:red, 204; green, 230; blue, 255 }  ,fill opacity=1 ] (264.67,133.27) .. controls (264.67,133.27) and (264.67,133.27) .. (264.67,133.27) -- (264.67,153.27) .. controls (264.67,153.27) and (264.67,153.27) .. (264.67,153.27) -- (204.67,153.27) .. controls (204.67,153.27) and (204.67,153.27) .. (204.67,153.27) -- (204.67,133.27) .. controls (204.67,133.27) and (204.67,133.27) .. (204.67,133.27) -- cycle ;
\draw    (215.07,116.95) .. controls (215.2,103.7) and (229.87,99.7) .. (230.2,84.37) ;
\draw    (230.4,110.28) .. controls (230.53,97.03) and (215.2,107.7) .. (215.53,92.37) ;
\draw [color={rgb, 255:red, 0; green, 0; blue, 0 }  ,draw opacity=0.27 ] [dash pattern={on 3pt off 1.5pt}]  (215.53,89.48) .. controls (212.87,77.73) and (180.53,74.43) .. (180.87,132.77) .. controls (181.2,191.1) and (211.87,191.84) .. (215.07,177.9) ;
\draw [color={rgb, 255:red, 0; green, 0; blue, 0 }  ,draw opacity=0.27 ] [dash pattern={on 3pt off 1.5pt}]  (230.2,80.67) .. controls (227.53,68.93) and (195.2,65.63) .. (195.53,123.96) .. controls (195.87,182.3) and (227.2,184.5) .. (230.4,170.56) ;
\draw [color={rgb, 255:red, 0; green, 0; blue, 0 }  ,draw opacity=0.27 ] [dash pattern={on 3pt off 1.5pt}]  (254.45,89.81) .. controls (257.32,78.07) and (292.22,74.77) .. (291.86,133.1) .. controls (291.5,191.44) and (258.4,192.17) .. (254.95,178.23) ;
\draw [color={rgb, 255:red, 0; green, 0; blue, 0 }  ,draw opacity=0.27 ] [dash pattern={on 3pt off 1.5pt}]  (270.28,80.34) .. controls (273.16,68.6) and (308.06,65.29) .. (307.7,123.63) .. controls (307.34,181.97) and (273.52,184.17) .. (270.07,170.23) ;
\draw  [fill={rgb, 255:red, 245; green, 166; blue, 35 }  ,fill opacity=1 ] (218.67,118.1) .. controls (218.67,118.1) and (218.67,118.1) .. (218.67,118.1) -- (218.67,125.1) .. controls (218.67,125.1) and (218.67,125.1) .. (218.67,125.1) -- (211.67,125.1) .. controls (211.67,125.1) and (211.67,125.1) .. (211.67,125.1) -- (211.67,118.1) .. controls (211.67,118.1) and (211.67,118.1) .. (211.67,118.1) -- cycle ;
\draw  [fill={rgb, 255:red, 245; green, 166; blue, 35 }  ,fill opacity=1 ] (234,112.1) .. controls (234,112.1) and (234,112.1) .. (234,112.1) -- (234,119.1) .. controls (234,119.1) and (234,119.1) .. (234,119.1) -- (227,119.1) .. controls (227,119.1) and (227,119.1) .. (227,119.1) -- (227,112.1) .. controls (227,112.1) and (227,112.1) .. (227,112.1) -- cycle ;

\end{mytikz4}\quad,
\end{equation}
where the orange blocks represent 
the projector $\Pi_q\vert_{A}$. Substitute the expectation $\mathbb{E}_\mathbb{U}[\rho^{\otimes 2}]$ and we obtain
\begin{equation}
\begin{aligned}
    \mathbb{E}_\mathbb{U} [{\rm tr}(\rho_{A,Q}^2)] &= \sum_q {\rm tr}\left[\mathbb{E}_\mathbb{U}[\rho^{\otimes 2}] ((\Pi_q\vert_{A})^{\otimes 2} S\vert_{2A}) \otimes I\vert_{2\bar{A}} \right] \\
    &= \sum_{q=0}^{|A|} {\rm tr}\left[ \frac{I+S}{d(d+1)} ((\Pi_q\vert_{A})^{\otimes 2} S\vert_{2A})\otimes I\vert_{2\bar{A}}) \right] \\
    &= \sum_{q=0}^{|A|} \frac{d_{\bar{A}}^2}{d(d+1)} {\rm tr}\left[ (\Pi_q\vert_{A})^2 \right] + \frac{d_{\bar{A}}}{d(d+1)} {\rm tr}\left[(\Pi_q\vert_{A})^{\otimes 2} \right] \\
    &= \frac{d_{\bar{A}}^2 }{d(d+1)} \sum_{q=0}^{|A|} \binom{|A|}{q} + \frac{d_{\bar{A}}}{d(d+1)} \sum_{q=0}^{|A|} \binom{|A|}{q}^2 \\
    &= \frac{d_{A} d_{\bar{A}}^2 }{d(d+1)} + \frac{d_{\bar{A}}}{d(d+1)} \binom{2|A|}{|A|} \\
    &= \frac{ 1 }{2^N+1} \left[ 2^{|\bar{A}|} + \frac{(2|A|)!}{2^{|A|}(|A|!)^2} \right],
\end{aligned}
\end{equation}
where we have used a special case of Vandermonde's identity, i.e., the summation of the squared binomial coefficients chosen from $n$ equals the central binomial coefficient chosen from $2n$. The ensemble average of the $2$-degree R\'enyi entanglement asymmetry, i.e., the difference between the $2$-degree R\'enyi entropies of the pruned state and the original state, can be approximated by the negative logarithm of the ratio of the two average purities above for the sake of the concentration of measure~\cite{ares2023entanglement}
\begin{equation}\label{eq:asymm_no_symmetry}
\begin{aligned}
    \Delta S_A^{(2)} &= S_2(\rho_{A,Q}) - S_2(\rho_{A}) \approx -\log \frac{\mathbb{E}_\mathbb{U} [{\rm tr}(\rho_{A,Q}^2)]}{\mathbb{E}_\mathbb{U} [{\rm tr}(\rho_{A}^2)]} \\
    &= -\log\left[ \frac{1}{1 + 2^{2|A|-N}} \left( 1 + \frac{(2|A|)!}{2^N (|A|!)^2} \right) \right] \\
    &\approx -\log\left[ \frac{1}{1 + 2^{2|A|-N}} \left( 1 + \frac{4^{|A|}}{2^N \sqrt{\pi |A|}} \right) \right] \\
    &= -\log\left[ \frac{1 + 4^{|A|-N/2}/\sqrt{\pi |A|} }{1 + 4^{|A|-N/2}} \right].
\end{aligned}
\end{equation}
One can see that in the large size limit $N\rightarrow\infty$, $ \mathbb{E} [\Delta S_A^{(2)}]$ is almost zero when $|A|<N/2$ and sharply changes to a non-zero value $\log\sqrt{\pi|A|}$ when $|A|>N/2$.

\subsection{Calculation with U(1) symmetry in the long-time limit}

Next, we consider the random circuits with the global U(1)-symmetry restriction. That is to say, every gate in the circuit should commute with the on-site symmetry operator $e^{-iZ_j\theta}$, i.e.,
\begin{equation}
\begin{mytikz2}
\draw    (169,128.68) -- (169,206.94) ;
\draw    (129.4,129.08) -- (129.4,206.94) ;
\draw  [fill={rgb, 255:red, 204; green, 230; blue, 255 }  ,fill opacity=1 ] (173,169) .. controls (176.31,169) and (179,171.69) .. (179,175) -- (179,183) .. controls (179,186.31) and (176.31,189) .. (173,189) -- (125,189) .. controls (121.69,189) and (119,186.31) .. (119,183) -- (119,175) .. controls (119,171.69) and (121.69,169) .. (125,169) -- cycle ;
\draw  [fill={rgb, 255:red, 255; green, 200; blue, 200 }  ,fill opacity=1 ] (121.8,148.1) .. controls (121.8,143.96) and (125.16,140.6) .. (129.3,140.6) .. controls (133.44,140.6) and (136.8,143.96) .. (136.8,148.1) .. controls (136.8,152.24) and (133.44,155.6) .. (129.3,155.6) .. controls (125.16,155.6) and (121.8,152.24) .. (121.8,148.1) -- cycle ;
\draw  [fill={rgb, 255:red, 255; green, 200; blue, 200 }  ,fill opacity=1 ] (161.4,148.1) .. controls (161.4,143.96) and (164.76,140.6) .. (168.9,140.6) .. controls (173.04,140.6) and (176.4,143.96) .. (176.4,148.1) .. controls (176.4,152.24) and (173.04,155.6) .. (168.9,155.6) .. controls (164.76,155.6) and (161.4,152.24) .. (161.4,148.1) -- cycle ;
\end{mytikz2}
\quad=\quad
\begin{mytikz2}
\draw    (270,109.88) -- (270,188.14) ;
\draw    (230.4,110.28) -- (230.4,188.14) ;
\draw  [fill={rgb, 255:red, 204; green, 230; blue, 255 }  ,fill opacity=1 ] (274,126.6) .. controls (277.31,126.6) and (280,129.29) .. (280,132.6) -- (280,140.6) .. controls (280,143.91) and (277.31,146.6) .. (274,146.6) -- (226,146.6) .. controls (222.69,146.6) and (220,143.91) .. (220,140.6) -- (220,132.6) .. controls (220,129.29) and (222.69,126.6) .. (226,126.6) -- cycle ;
\draw  [fill={rgb, 255:red, 255; green, 200; blue, 200 }  ,fill opacity=1 ] (223.2,168.1) .. controls (223.2,163.96) and (226.56,160.6) .. (230.7,160.6) .. controls (234.84,160.6) and (238.2,163.96) .. (238.2,168.1) .. controls (238.2,172.24) and (234.84,175.6) .. (230.7,175.6) .. controls (226.56,175.6) and (223.2,172.24) .. (223.2,168.1) -- cycle ;
\draw  [fill={rgb, 255:red, 255; green, 200; blue, 200 }  ,fill opacity=1 ] (262.8,168.1) .. controls (262.8,163.96) and (266.16,160.6) .. (270.3,160.6) .. controls (274.44,160.6) and (277.8,163.96) .. (277.8,168.1) .. controls (277.8,172.24) and (274.44,175.6) .. (270.3,175.6) .. controls (266.16,175.6) and (262.8,172.24) .. (262.8,168.1) -- cycle ;
\end{mytikz2}\quad,
\end{equation}
where the blue square represents the given gate and the red circles represent the Pauli-Z rotation gates. To construct the ensemble $\mathbb{U}$ that respects a certain symmetry, one reasonable choice is to assemble the Haar measures over each subspace of irreducible representation (irrep) for that symmetry. That is to say, under the eigenbasis of the symmetry, we require each sub-block to be an independent Haar-random unitary, i.e.,
\begin{equation}\label{eq:general_symm_haar}
    \mathbb{U} = \left\{U \mid U = \bigoplus_q  (U_q  \otimes I_{ q }),~U_q \text{ is Haar-random} \right\},
\end{equation}
where $q$ is the label of irreps. $I_q$ is the identity of dimension $d_q $ in a single irrep of $q$. $U_q$ is a unitary of dimension $r_q$ on the tensor-product-quotient space of the $r_q$ repeated irreps of $q$. The symmetric ensemble $\mathbb{U}$ is defined by taking each $U_q$ as an independent Haar-random matrix.

For the global U(1) symmetry, $q$ just corresponds to the total charge number $\hat{Q}=\sum_i n_i$ or say the total spin $z$-component $\sum_i \sigma^z_i$. As the U(1) group is Abelian, its irreps are all one-dimensional $d_q =1$, which are just the phase factors with different frequencies $e^{i q \theta}$. The number of repeated irreps of $q$ equals the number of occupation configurations corresponding to $q$ charges on $N$ sites, i.e., $r_q  = \binom{N}{ q }$. For simplicity, we replace the direct sum by a normal sum with projectors 
\begin{equation}\label{eq:u1_haar_def}
    \mathbb{U} = \left\{U \mid U = \sum_q  U_q  \Pi_q  \right\},
\end{equation}
where $\Pi_q $ is the projector onto the subspace of $ q $, i.e., the $q$-charge sector. Note that $\Pi_q $ corresponds to the subspace of all the repeated irreps of $q$, instead of just one of them. Compared to Eq.~\eqref{eq:general_symm_haar}, $U_q$ in Eq.~\eqref{eq:u1_haar_def} is enlarged to the whole Hilbert space by padding identities. Under the eigenbasis of the symmetry, $U_q$ and $\Pi_q$ take the form of
\begin{equation}
    U_q  = \left(\begin{matrix}
        * & 0 \\
        0 & I
    \end{matrix}\right),\quad   \Pi_q  = \left(\begin{matrix}
        I & 0 \\
        0 & 0
    \end{matrix}\right),
\end{equation}
where the upper left block represents the $q$-charge sector and the lower right block represents the others. Hence we naturally have $U_q  \Pi_q  = \Pi_q  U_q $. In other words, $U_q  \Pi_q $ can be seen as a pseudo-unitary padded with zeros outside the $q$-charge sector. We will denote 
\begin{equation}
    V_q =U_q  \Pi_q
\end{equation}
for simplicity below. 

We start from the first-order integration. Since Haar averaging requires that each unitary is paired with its conjugation (otherwise the integral vanishes), the first-order integration over the U(1)-symmetric ensemble $\mathbb{U}$ becomes
\begin{equation}\label{eq:u1_twirl_t1}
\begin{aligned}
    \mathbb{E}_{\mathbb{U}}\left[ \rho \right] &= \mathbb{E}_{\mathbb{U}}\left[ U^\dg \rho_0 U \right] = \mathbb{E}_{\mathbb{U}}\left[ \left(\sum_q  V_q ^\dg \right) \rho_0 \left(\sum_{ q'} V_{ q'}\right) \right] \\
    &= \sum_{ q  q'}\mathbb{E}_{\mathbb{U}}\left[ V_q ^\dg \rho_0 V_{ q'} \right] = \sum_{ q }\mathbb{E}_{\mathbb{U}}\left[ V_q ^\dg \rho_0 V_{ q } \right] \\
    &= \sum_{ q } \mathbb{E}_{\mathbb{U}}\left[ U_q ^\dg \Pi_q  \rho_0 \Pi_q   U_{ q } \right] = \sum_q  \frac{{\rm tr}(\Pi_q  \rho_0 \Pi_q )}{r_q } \Pi_q = \sum_q  \frac{{\rm tr}(\rho_0 \Pi_q )}{r_q } \Pi_q .
\end{aligned}
\end{equation}
One can see that the result is almost the maximally mixed state but with different weights on different charge sectors, which respects the weak symmetry $e^{i\hat{Q}\theta} \mathbb{E}_{\mathbb{U}}\left[ \rho \right] e^{-i\hat{Q}\theta} = \mathbb{E}_{\mathbb{U}}\left[ \rho \right]$. Moreover, if the initial state $\rho_0$ respects the strong symmetry $e^{i\hat{Q}\theta}\rho_0=e^{i q \theta}\rho_0$, i.e., the non-zero elements in $\rho_0$ is within a single charge sector, then Eq.~\eqref{eq:u1_twirl_t1} tells us that the evolved state will be the ``maximally mixed state'' within the $q$-charge sector, which still respects the strong symmetry, exhibiting the charge conservation law of the evolution. 

After partially tracing out subsystem $\bar{A}$, the result is still a diagonal matrix under the computational basis, implying that the averaged reduced density matrix $\mathbb{E}_\mathbb{U}[\rho_A]$ must also respect the weak symmetry. To be specific, by definition, we know the projector $\Pi_q$ can be decomposed as
\begin{equation}\label{eq:decompose_proj_q}
    \Pi_q  = \sum_{ q'=0}^{q} \Pi_{ q'}\vert_{A} \otimes \Pi_{ q  -  q'}\vert_{\bar{A}},
\end{equation}
where $\Pi_{ q'}\vert_{A}$ denotes the projector to the $ q'$ charge subspace restricted on the subsystem $A$. By default, we assume if $q'>|A|$ then $\Pi_{q'}\vert_{A}=0$. Hence the index $q'$ actually takes values from $\max\{0,q-|\bar{A}|\}$ to $\min\{q,|A|\}$ in Eq.~\eqref{eq:decompose_proj_q}. The partial trace of the projector is
\begin{equation}\label{eq:u1_rho_A}
\begin{aligned}
    &{\rm tr}_{\bar{A}}\Pi_{ q } = \sum_{ q'} \Pi_{ q'}\vert_{A} {\rm tr}(\Pi_{ q  -  q'}\vert_{\bar{A}}) = \sum_{ q'=0}^{q} \binom{|\bar{A}|}{q-q'} \Pi_{ q'}\vert_{A}.
\end{aligned}
\end{equation}
We denote $\binom{n}{k}=0$ if $k>n$. Hence, the averaged reduced density matrix is
\begin{equation}
    \mathbb{E}_\mathbb{U} [\rho_A] = \mathbb{E}_\mathbb{U} [{\rm tr}_{\bar{A}} \rho] = \sum_q  \frac{{\rm tr}(\Pi_q  \rho_0 \Pi_q )}{r_q } {\rm tr}_{\bar{A}} \Pi_q  = \sum_{ q' q } \frac{{\rm tr}(\rho_0 \Pi_q )}{r_q } \binom{|\bar{A}|}{ q  -  q'} \Pi_{ q'}\vert_{A},
\label{eq:u1rhog}
\end{equation}
One can see that the result depends on the initial state $\rho_0$ only via the factor ${\rm tr}(\rho_0\Pi_q)$. If the initial state takes the form of the $Y$-tilted ferromagnetic state
\begin{equation}
    e^{-\frac{i}{2}\sum_j \sigma^{y}_j\theta} \ket{0}^{\otimes N} = \left(\cos\frac{\theta}{2}\ket{0} + \sin\frac{\theta}{2}\ket{1}\right)^{\otimes N} = \sum_b \left(\cos\frac{\theta}{2}\right)^{N-q(b)} \left(\sin\frac{\theta}{2}\right)^{q(b)} \ket{b},
\end{equation}
where $b$ represents $01$-bit strings and $q(b)$ counts the number of $1$ in $b$, then the weight in the $q$-charge sector is
\begin{equation}
\begin{aligned}
    {\rm tr}(\rho_0 \Pi_q) &= \sum_b \left(\cos^2\frac{\theta}{2}\right)^{N-q(b)} \left(\sin^2\frac{\theta}{2}\right)^{q(b)} \braoprket{b}{\Pi_q}{b} \\
    &= \sum_b \left(\cos^2\frac{\theta}{2}\right)^{N-q(b)} \left(\sin^2\frac{\theta}{2}\right)^{q(b)} \delta_{q(b),q} \\
    &= r_q \left(\cos^2\frac{\theta}{2}\right)^{N-q} \left(\sin^2\frac{\theta}{2}\right)^{q}.
\end{aligned}
\end{equation}
The averaged reduced density matrix hence becomes
\begin{equation}
\begin{aligned}
    \mathbb{E}_\mathbb{U} [\rho_A] &= \sum_{q'=0}^{|A|} \sum_{ q=q' }^{q'+|\bar{A}|} \left(\cos^2\frac{\theta}{2}\right)^{N-q} \left(\sin^2\frac{\theta}{2}\right)^{q} \binom{|\bar{A}|}{ q  -  q'} \Pi_{ q'}\vert_{A} \\
    &= \left(\cos^2\frac{\theta}{2}\right)^{|A|} \sum_{q'=0}^{|A|} \left(\tan^2\frac{\theta}{2}\right)^{q'} \Pi_{ q'}\vert_{A} ,
\end{aligned}
\end{equation}
where the summation over $(q-q')$ is simply the expansion of the binomial $\left(\cos^2\frac{\theta}{2}+\sin^2\frac{\theta}{2}\right)^{|\bar{A}|}$. One can see that when $\theta\rightarrow 0$ or $\theta\rightarrow \pi$, the weights concentrate on the $q'=0$ or $q'=|A|$ charge sector, consistent with the fact that the initial all-zero / all-one state will not be changed and thus not thermalize by the U(1)-symmetric evolution. When $\theta\in(0,\pi/2)$, the weights form an exponential distribution decaying from $q'=0$ to $q'=|A|$ with the decaying rate $-\log\left(\tan^2\frac{\theta}{2}\right)$. Similarly, when $\theta\in(\pi/2,\pi)$, the weights also form an exponential distribution but decaying from $q'=|A|$ to $q'=0$. If $\theta=\pi/2$, the weights are uniform, i.e., the averaged reduced density matrix becomes exactly the maximally mixed state. It is worth noticing that the results depend on the initial state $\rho_0$ only through the initial distribution over different charge sectors, i.e., the factor ${\rm tr}(\rho_0 \Pi_q)$. The same is true for the second-order integration below.

\begin{figure}
    \centering
    \includegraphics[width=0.45\linewidth]{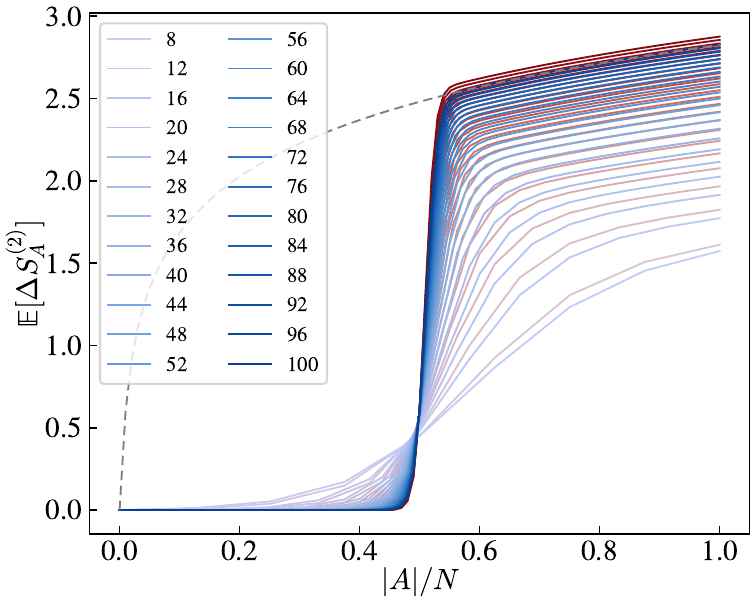}
    \caption{The average R\'enyi-2 entanglement asymmetry $\mathbb{E}[\Delta S^{(2)}_A]$ in the long-time limit of U(1)-symmetric random circuit evolution with the tilt angle $\theta=0.4\pi$ from the ferromagnetic initial state versus the subsystem size $|A|$. The numbers in the legend represent different system sizes $N$. The blue and red lines represent the results from U(1)-symmetric circuits and non-symmetric circuits, respectively. The grey dashed line represents the result for the tilted ferromagnetic initial state with $N=100$. }
    \label{fig:deltaS_A_04pi}
\end{figure}

Next, we consider the second-order integration. According to the conjugation-pair rule of Haar averaging, the second-order integration can be expanded as
\begin{equation}\label{eq:u1_twirl_t2}
\begin{aligned}
    \mathbb{E}_{\mathbb{U}}\left[ U^{\dg \otimes 2} A U^{\otimes 2} \right] &= \mathbb{E}_{\mathbb{U}}\left[ \left(\sum_q  V_q ^\dg \right)^{ \otimes 2} A \left(\sum_{ q' } V_{ q' } \right)^{\otimes 2} \right] = \sum_{ q p q' p'}\mathbb{E}_{\mathbb{U}}\left[ (V_q ^\dg \otimes V_p^\dg) A (V_{ q'}\otimes V_{p'}) \right] \\
    &= \sum_{ q \neq p}\mathbb{E}_{\mathbb{U}}\left[ (V_q ^\dg \otimes V_p^\dg) A (V_{ q }\otimes V_{p}) \right] + \mathbb{E}_{\mathbb{U}}\left[ (V_q ^\dg \otimes V_p^\dg) A (V_{p}\otimes V_{ q }) \right] \\
    & \quad+ \sum_{ q }\mathbb{E}_{\mathbb{U}}\left[ (V_q ^\dg \otimes V_q ^\dg) A (V_{ q }\otimes V_{ q }) \right].
\end{aligned}
\end{equation}
The first term can be integrated as
\begin{equation}
\begin{aligned}
    \sum_{ q \neq p}\mathbb{E}_{\mathbb{U}}\left[ (V_q ^\dg \otimes V_p^\dg) A (V_{ q }\otimes V_{p}) \right] = \sum_{ q \neq p} \frac{{\rm tr}[A (\Pi_q  \otimes \Pi_p)]}{r_q  r_p} \Pi_q  \otimes \Pi_p.
\end{aligned}
\end{equation}
The second term can be integrated as
\begin{equation}
\begin{aligned}
    \sum_{ q \neq p}\mathbb{E}_{\mathbb{U}}\left[ (V_q ^\dg \otimes V_p^\dg) A (V_{p}\otimes V_{ q }) \right] = \sum_{ q \neq p} \frac{{\rm tr}[A S (\Pi_q  \otimes \Pi_p)]}{r_q  r_p} (\Pi_q  \otimes \Pi_p) S,
\end{aligned}
\end{equation}
where $S$ is the SWAP operator on the two replicas. The third term can be integrated as
\begin{equation}
\begin{aligned}
    &\sum_{ q }\mathbb{E}_{\mathbb{U}}\left[ (V_q ^\dg \otimes V_q ^\dg) A (V_{ q }\otimes V_{ q }) \right] \\
    &= \sum_q  \left[\frac{{\rm tr}(A \Pi_q ^{\otimes 2}) I + {\rm tr}(SA \Pi_q ^{\otimes 2}) S}{r_q ^2-1} - \frac{{\rm tr}(A \Pi_q ^{\otimes 2}) S + {\rm tr}(SA \Pi_q ^{\otimes 2}) I}{r_q (r_q ^2-1)}\right] \Pi_q ^{\otimes 2}.
\end{aligned}
\end{equation}
If we take $A=\rho_0^{\otimes 2}$, then the above three terms in $\mathbb{E}_\mathbb{U}[\rho^{\otimes 2}]$ reduce to
\begin{equation}
    \sum_{ q \neq p}\mathbb{E}_{\mathbb{U}}\left[ (V_q ^\dg \otimes V_p^\dg) \rho_0^{\otimes 2} (V_{ q }\otimes V_{p}) \right] = \sum_{ q \neq p} \frac{{\rm tr}(\rho_0 \Pi_q ) {\rm tr}(\rho_0 \Pi_p)}{r_q  r_p} \Pi_q  \otimes \Pi_p,
\end{equation}
\begin{equation}
\begin{aligned}
    \sum_{ q \neq p}\mathbb{E}_{\mathbb{U}}\left[ (V_q ^\dg \otimes V_p^\dg) \rho_0^{\otimes 2} (V_{p}\otimes V_{ q }) \right] &= \sum_{ q \neq p} \frac{{\rm tr}(\rho_0 \Pi_q  \rho_0 \Pi_p)}{r_q  r_p} (\Pi_q  \otimes \Pi_p)S \\
    &= \sum_{ q \neq p} \frac{{\rm tr}(\rho_0 \Pi_q  ) {\rm tr}(\rho_0 \Pi_p)}{r_q  r_p} (\Pi_q  \otimes \Pi_p)S, 
\end{aligned}
\end{equation}
\begin{equation}
\begin{aligned}
    & \sum_{ q }\mathbb{E}_{\mathbb{U}}\left[ (V_q ^\dg \otimes V_q ^\dg) \rho_0^{\otimes 2} (V_{ q }\otimes V_{ q }) \right] \\
    & = \sum_q  \left[\frac{{\rm tr}(\rho_0 \Pi_q )^2 I + {\rm tr}((\rho_0 \Pi_q )^2) S}{r_q ^2-1} - \frac{{\rm tr}(\rho_0 \Pi_q )^2 S + {\rm tr}((\rho_0 \Pi_q )^2) I}{r_q (r_q ^2-1)}\right] \Pi_q ^{\otimes 2} \\
    & = \sum_q  \frac{ {\rm tr}(\rho_0 \Pi_q )^2 }{r_q (r_q +1)} (I+S) \Pi_q ^{\otimes 2},
\end{aligned}
\end{equation}
where we have used the fact that $\rho_0$ is a pure state so that ${\rm tr}(\rho_0^2)=1$ and
\begin{equation}
\begin{aligned}
    & {\rm tr}(\rho_0 \Pi_q  \rho_0 \Pi_p)={\rm tr}(\rho_0 \Pi_q ) {\rm tr}(\rho_0 \Pi_p), \\
    & {\rm tr}((\rho_0 \Pi_q )^2) = ({\rm tr}(\rho_0 \Pi_q ))^2.
\end{aligned}
\end{equation}
Hence, the expectation of the purity of $\rho_A$ is
\begin{equation}\label{eq:purity_expectation_u1}
\begin{aligned}
    &\mathbb{E}_\mathbb{U} [{\rm tr}(\rho_A^2)] = {\rm tr}\left[\mathbb{E}_\mathbb{U}[\rho^{\otimes 2}] (S\vert_{2A}\otimes I\vert_{2\bar{A}})\right] \\
    &= \sum_{ q \neq p} \frac{{\rm tr}(\rho_0 \Pi_q ) {\rm tr}(\rho_0 \Pi_p)}{r_q  r_p} \left[ {\rm tr}({\rm tr}_{\bar{A}}\Pi_{ q } {\rm tr}_{\bar{A}}\Pi_{p}) + {\rm tr}({\rm tr}_{A}\Pi_{ q } {\rm tr}_{A}\Pi_{p}) \right] \\
    &\quad + \sum_q  \frac{ {\rm tr}(\rho_0 \Pi_q )^2 }{r_q (r_q +1)}  \left[ {\rm tr}(({\rm tr}_{\bar{A}}\Pi_{ q })^2) + {\rm tr}(({\rm tr}_{A}\Pi_{ q })^2) \right].
\end{aligned}
\end{equation}
The involved trace factors can be calculated as
\begin{equation}\label{eq:f_func_def}
\begin{aligned}
    {\rm tr}({\rm tr}_{\bar{A}}\Pi_{ q } {\rm tr}_{\bar{A}}\Pi_{p}) &= \sum_{q'p'} \binom{|\bar{A}|}{q-q'} \binom{|\bar{A}|}{p-p'} {\rm tr}(\Pi_{ q'}\vert_{A} \Pi_{ p'}\vert_{A}) \\
    &= \sum_{q'} \binom{|\bar{A}|}{q-q'} \binom{|\bar{A}|}{p-q'}  \binom{|A|}{q'} \\
    &= \sum_{q'=0}^{\min\{q,p\}} \binom{|\bar{A}|}{q-q'} \binom{|\bar{A}|}{p-q'}  \binom{|A|}{q'} \\
    & \equiv f(q,p,|A|,|\bar{A}|),
\end{aligned}
\end{equation}
\begin{equation}
\begin{aligned}
    {\rm tr}({\rm tr}_{A}\Pi_{ q } {\rm tr}_{A}\Pi_{p}) &= f(q,p,|\bar{A}|,|A|),
\end{aligned}
\end{equation}
\begin{equation}
\begin{aligned}
    {\rm tr}(({\rm tr}_{\bar{A}}\Pi_{ q })^2 ) &= f(q,q,|A|,|\bar{A}|),
\end{aligned}
\end{equation}
\begin{equation}
\begin{aligned}
    {\rm tr}(({\rm tr}_{A}\Pi_{ q })^2 ) &= f(q,q,|\bar{A}|,|A|),
\end{aligned}
\end{equation}
Remember that $\binom{n}{k}=0$ if $k>n$ and hence the index $q'$ actually takes values from $\max\{0,q-|\bar{A}|,p-|\bar{A}|\}$ to $\min\{q,p,|A|\}$ in the definition of $f(q,p,|A|,|\bar{A}|)$ in Eq.~\eqref{eq:f_func_def}. If the initial state takes the form of the $Y$-tilted ferromagnetic state, the two coefficients in Eq.~\eqref{eq:purity_expectation_u1} become
\begin{equation}
\begin{aligned}
    & \frac{{\rm tr}(\rho_0 \Pi_q ) {\rm tr}(\rho_0 \Pi_p)}{r_q  r_p} = \left(\cos^2\frac{\theta}{2}\right)^{2N-q-p} \left(\sin^2\frac{\theta}{2}\right)^{q+p}, \\
    & \frac{ {\rm tr}(\rho_0 \Pi_q )^2 }{r_q (r_q +1)} = \frac{r_q}{r_q+1} \left(\cos^2\frac{\theta}{2}\right)^{2N-2q} \left(\sin^2\frac{\theta}{2}\right)^{2q}.
\end{aligned}
\end{equation}
Thus, the average purity of $\rho_A$ becomes 
\begin{equation}\label{eq:u1_purity_rho_A}
\begin{aligned}
    &\mathbb{E}_\mathbb{U} [{\rm tr}(\rho_A^2)] = \sum_{ q\neq p} \left(\cos^2\frac{\theta}{2}\right)^{2N-q-p} \left(\sin^2\frac{\theta}{2}\right)^{q+p} [f(q,p,|A|,|\bar{A}|)+f(q,p,|\bar{A}|,|A|)]  \\
    &\quad + \sum_q  \frac{r_q}{r_q+1} \left(\cos^2\frac{\theta}{2}\right)^{2N-2q} \left(\sin^2\frac{\theta}{2}\right)^{2q} \left[ f(q,q,|A|,|\bar{A}|) + f(q,q,|\bar{A}|,|A|) \right].
\end{aligned}
\end{equation}
On the other hand, the purity of the pruned reduced state is
\begin{equation}
\begin{aligned}
    {\rm tr}(\rho_{A,Q}^2) &= {\rm tr}\left( \sum_{qp} \Pi_q\vert_{A} \rho_A \Pi_q\vert_{A}  \Pi_p\vert_{A} \rho_A \Pi_p\vert_{A} \right) \\
    &= \sum_{q} {\rm tr}\left( \rho_A \Pi_q\vert_{A} \rho_A \Pi_q\vert_{A} \right) \\
    &= \sum_{q} {\rm tr}\left[ \rho^{\otimes 2} ((\Pi_q\vert_{A})^{\otimes 2} S\vert_{2A}) \otimes I_{2\bar{A}} \right] .
\end{aligned}
\end{equation}
Thus the average purity of the pruned state is
\begin{equation}
\begin{aligned}
    \mathbb{E}_\mathbb{U} [{\rm tr}(\rho_{A,Q}^2)] &= \sum_{q'} {\rm tr}\left[\mathbb{E}_\mathbb{U}[\rho^{\otimes 2}] ((\Pi_{q'}\vert_{A})^{\otimes 2} S\vert_{2A}) \otimes I\vert_{2\bar{A}} \right] \\
    &= \sum_{ q \neq p, q'} \frac{{\rm tr}(\rho_0 \Pi_q ) {\rm tr}(\rho_0 \Pi_p)}{r_q  r_p} \bigg[ {\rm tr}\big( ({\rm tr}_{\bar{A}}\Pi_{ q }) (\Pi_{q'}\vert_{A}) ({\rm tr}_{\bar{A}}\Pi_{p}) (\Pi_{q'}\vert_{A})\big) \\
    &\quad\quad\quad\quad\quad\quad + {\rm tr}({\rm tr}_{A}(\Pi_{ q }(\Pi_{q'}\vert_{A})) {\rm tr}_{A}(\Pi_{p}(\Pi_{q'}\vert_{A}))) \bigg]  \\
    &\quad + \sum_{q,q'}  \frac{ {\rm tr}(\rho_0 \Pi_q )^2 }{r_q (r_q +1)}  \left[ {\rm tr}(( ({\rm tr}_{\bar{A}}\Pi_{ q }) (\Pi_{q'}\vert_{A}))^2) + {\rm tr}(({\rm tr}_{A}(\Pi_{ q }(\Pi_{q'}\vert_{A}) ))^2) \right].
\end{aligned}
\end{equation}
The subterms take the form of
\begin{equation}
\begin{aligned}
     {\rm tr}\big( ({\rm tr}_{\bar{A}}\Pi_{ q }) (\Pi_{q'}\vert_{A}) ({\rm tr}_{\bar{A}}\Pi_{p}) (\Pi_{q'}\vert_{A})\big) &= \binom{|\bar{A}|}{q-q'} \binom{|\bar{A}|}{p-q'}  \binom{|A|}{q'},
\end{aligned}
\end{equation}

\begin{equation}
\begin{aligned}
    {\rm tr}({\rm tr}_{A}(\Pi_{ q }(\Pi_{q'}\vert_{A})) {\rm tr}_{A}(\Pi_{p}(\Pi_{q'}\vert_{A}))) &= \binom{|A|}{q'} \binom{|A|}{q'}  \binom{|\bar{A}|}{q-q'}\delta_{pq},
\end{aligned}
\end{equation}

\begin{equation}
\begin{aligned}
    {\rm tr}(( ({\rm tr}_{\bar{A}}\Pi_{ q }) (\Pi_{q'}\vert_{A}))^2) &= \binom{|\bar{A}|}{q-q'} \binom{|\bar{A}|}{q-q'}  \binom{|A|}{q'},
\end{aligned}
\end{equation}

\begin{equation}
\begin{aligned}
    {\rm tr}(({\rm tr}_{A}(\Pi_{ q }(\Pi_{q'}\vert_{A}) ))^2) &= \binom{|A|}{q'} \binom{|A|}{q'}  \binom{|\bar{A}|}{q-q'}.
\end{aligned}
\end{equation}
Hence, the average purity of $\rho_{A,Q}$ becomes
\begin{equation}\label{eq:u1_purity_rho_AQ}
\begin{aligned}
    &\mathbb{E}_\mathbb{U} [{\rm tr}(\rho_{A,Q}^2)] = \sum_{ q\neq p} \left(\cos^2\frac{\theta}{2}\right)^{2N-q-p} \left(\sin^2\frac{\theta}{2}\right)^{q+p} f(q,p,|A|,|\bar{A}|)  \\
    &\quad + \sum_q  \frac{r_q}{r_q+1} \left(\cos^2\frac{\theta}{2}\right)^{2N-2q} \left(\sin^2\frac{\theta}{2}\right)^{2q} \left[ f(q,q,|A|,|\bar{A}|) + f(q,q,|\bar{A}|,|A|) \right].
\end{aligned}
\end{equation}
One can see that Eq.~\eqref{eq:u1_purity_rho_AQ} is almost the same as Eq.~\eqref{eq:u1_purity_rho_A}, except missing the term involving $f(q,p,|\bar{A}|,|A|)$. The R\'enyi entanglement asymmetry can be obtained by substituting Eqs.~\eqref{eq:u1_purity_rho_A} and \eqref{eq:u1_purity_rho_AQ} into
\begin{equation}
    \Delta S_A^{(2)} = S_2(\rho_{A,Q}) - S_2(\rho_{A}) \approx -\log \frac{\mathbb{E}_\mathbb{U} [{\rm tr}(\rho_{A,Q}^2)]}{\mathbb{E}_\mathbb{U} [{\rm tr}(\rho_{A}^2)]}.
\end{equation}
According to Eqs.~\eqref{eq:u1_purity_rho_A} and \eqref{eq:u1_purity_rho_AQ}, both $\mathbb{E}_\mathbb{U} [{\rm tr}(\rho_{A, Q}^2)]$ and $\mathbb{E}_\mathbb{U} [{\rm tr}(\rho_{A}^2)]$ are just sums of a polynomial number of certain binomial coefficients, which can be easily and accurately calculated numerically. But unfortunately, the symbolic summation of these binomial coefficients will lead to generalized hypergeometric functions, which are hard to write in an explicit closed form. However, in the large-size limit $N\rightarrow\infty$, the distribution of binomial coefficients converges to the Gaussian distribution and the summations can be well approximated by continuum integrals under certain conditions, which can give rise to a simple and meaningful analytical form of $ \mathbb{E} [\Delta S_A^{(2)}]$. To be specific, we first regard $\frac{r_q}{r_q+1}\approx 1$ (recall $r_q=\binom{N}{q}$). This makes $\mathbb{E}_\mathbb{U} [{\rm tr}(\rho_{A, Q}^2)]$ and $\mathbb{E}_\mathbb{U} [{\rm tr}(\rho_{A}^2)]$ reduce to
\begin{equation}\label{eq:u1_rho_A_tan}
\begin{aligned}
    \mathbb{E}_\mathbb{U} [{\rm tr}(\rho_{A}^2)] &\approx \left(\cos^2\frac{\theta}{2}\right)^{2N} \sum_{ q,p} \left(\tan^2\frac{\theta}{2}\right)^{q+p} [f(q,p,|A|,|\bar{A}|)+f(q,p,|\bar{A}|,|A|)].
\end{aligned}
\end{equation}
\begin{equation}\label{eq:u1_rho_AQ_tan}
\begin{aligned}
    \mathbb{E}_\mathbb{U} [{\rm tr}(\rho_{A,Q}^2)] &\approx \left(\cos^2\frac{\theta}{2}\right)^{2N} \sum_{ q,p} \left(\tan^2\frac{\theta}{2}\right)^{q+p} [f(q,p,|A|,|\bar{A}|)+f(q,p,|\bar{A}|,|A|)\delta_{qp}].
\end{aligned}
\end{equation}
One can find that up to this approximation, the only difference between the two terms is the Kronecker factor $\delta_{qp}$. Suppose $|A|$ and $|\bar{A}|$ grow linearly with $N$ in the large size limit. We use the Gaussian distribution to approximate the binomial coefficients
\begin{equation}
    \binom{n}{k} \approx \frac{2^n}{\sqrt{\pi n/2}} \exp\left[-\frac{(k-n/2)^2}{n}\right].
\end{equation}
Then the function $f(q,p,|A|,|\bar{A}|)$ can be approximated by
\begin{equation}
\begin{aligned}
    &f(q,p,|A|,|\bar{A}|) = \sum_{q'=0}^{\min\{q,p\}} \binom{|\bar{A}|}{q-q'} \binom{|\bar{A}|}{p-q'}  \binom{|A|}{q'} \\ 
    &\approx \frac{2^{2|\bar{A}|+|A|}}{\sqrt{\pi^3 |\bar{A}|^2 |A|/8}} \int dq' \exp\left[-\frac{(q-q'-|\bar{A}|/2)^2}{|\bar{A}|} -\frac{(p-q'-|\bar{A}|/2)^2}{|\bar{A}|} -\frac{(q'-|A|/2)^2}{|A|}\right] \\
    &= \frac{2^{2|\bar{A}|+|A|}}{\sqrt{\pi^3 |\bar{A}|^2 |A|/8}} \sqrt{\frac{\pi |A||\bar{A}|}{2|A|+|\bar{A}|}} \times \\
    &\quad\exp\left[ -\frac{|A|^2 |\bar{A}| + 2 |A| \left(|\bar{A}|^2 + (p - q)^2 - |\bar{A}| (p + q) \right) + |\bar{A}| \left(|\bar{A}|^2 - 2 |\bar{A}| (p + q) + 2 (p^2 + q^2) \right) }{2|\bar{A}|(2|A|+|\bar{A}|)} \right] \\
    &\equiv \tilde{f}(q,p,|A|,|\bar{A}|).
\end{aligned}
\end{equation}
By use of Gaussian integration, the two summations can be approximated by
\begin{equation}\label{eq:binom_to_gaussian_pq}
\begin{aligned}
     &\sum_{ q,p} \left(\tan^2\frac{\theta}{2}\right)^{q+p} f(q,p,|A|,|\bar{A}|) \approx \int dq dp \left(\tan^2\frac{\theta}{2}\right)^{q+p} \tilde{f}(q,p,|A|,|\bar{A}|) \\
     &= \frac{2^{2|\bar{A}|+|A|}}{\sqrt{\pi^3 |\bar{A}|^2 |A|/8}}\sqrt{\pi^3 |A||\bar{A}|^2} \exp\left[ 2 (2 |A| + |\bar{A}|) \log^2\left|\tan\frac{\theta}{2}\right|\right] \left(\tan^2\frac{\theta}{2}\right)^{2(|A| + |\bar{A}|)} \\
     &= 2^{N+|\bar{A}|} \sqrt{8} \exp\left[ 2 (N+ |A|) \log^2\left|\tan\frac{\theta}{2}\right|\right] \left(\tan^2\frac{\theta}{2}\right)^{N},
\end{aligned}
\end{equation}
and
\begin{equation}\label{eq:binom_to_gaussian_qq}
\begin{aligned}
     &\sum_{ q} \left(\tan^2\frac{\theta}{2}\right)^{2q} f(q,q,|A|,|\bar{A}|) \approx \int (\sqrt{2} dq) \left(\tan^2\frac{\theta}{2}\right)^{2q} \tilde{f}(q,q,|A|,|\bar{A}|) \\
     &= \frac{2^{2|\bar{A}|+|A|}}{\sqrt{\pi^3 |\bar{A}|^2 |A|/8}} \sqrt{2} \sqrt{\pi^2 |A||\bar{A}|/2} \exp\left[ 2 (2 |A| + |\bar{A}|) \log^2\left|\tan\frac{\theta}{2}\right|\right] \left(\tan^2\frac{\theta}{2}\right)^{2(|A| + |\bar{A}|)} \\
     &= 2^{N+|\bar{A}|} \sqrt{\frac{8}{\pi |\bar{A}|}} \exp\left[ 2 (N + |A|) \log^2\left|\tan\frac{\theta}{2}\right|\right] \left(\tan^2\frac{\theta}{2}\right)^{N},
\end{aligned}
\end{equation}
where the $\sqrt{2}$ factor in the integration measure in Eq.~\eqref{eq:binom_to_gaussian_qq} arises because the integral domain of Eq.~\eqref{eq:binom_to_gaussian_qq} is the diagonal line on the $p$-$q$ plane compared to Eq.~\eqref{eq:binom_to_gaussian_pq}. To make the integration results above valid and meaningful, they should satisfy $\mathbb{E}_\mathbb{U} [{\rm tr}(\rho_A^2)]\leq 1$ and $\mathbb{E}_\mathbb{U} [{\rm tr}(\rho_{A, Q}^2)]\leq 1$ especially for $N\rightarrow \infty$, which will give rise to extra restrictions on $\theta$. For example, if we take $|A|=|\bar{A}|=N/2$, then the restriction is
\begin{equation}
    2^{3/2}\times \exp\left[ 3 \log^2\left|\tan\frac{\theta}{2}\right|\right] \left(\cos^2\frac{\theta}{2}\right) \left(\sin^2\frac{\theta}{2}\right) \leq 1 \Rightarrow \frac{\pi}{2}-0.128\pi\leq \theta\leq \frac{\pi}{2}+0.128\pi,
\end{equation}
which means that this approximation can only be valid around $\theta=\frac{\pi}{2}$. This can be understood by the fact that when $\theta$ is too small, the factor $(\tan^2\frac{\theta}{2})^{q+p}$ in Eqs.~\eqref{eq:u1_rho_A_tan} and \eqref{eq:u1_rho_AQ_tan} will concentrate at the margin $q+p=0$, where the error of the Gaussian approximation is relatively large. Moreover, if one wants the approximated purity to be valid for arbitrary values of $|A|$, the extreme restriction will occur at $|\bar{A}|=0$ or $|A|=0$, i.e.,
\begin{equation}
    4\times \exp\left[ 2 \log^2\left|\tan\frac{\theta}{2}\right|\right] \left(\cos^2\frac{\theta}{2}\right) \left(\sin^2\frac{\theta}{2}\right) \leq 1 \Rightarrow \theta= \frac{\pi}{2}.
\end{equation}
Namely the Gaussian approximation is valid only at a single point $\theta=\frac{\pi}{2}$ if we consider arbitrary values of $|A|$ simultaneously. However, we will see below that the final result is actually reasonable in a finite region of $\theta$ around $\frac{\pi}{2}$. Substituting the integration results, the R\'enyi-2 entanglement asymmetry becomes
\begin{equation}\label{eq:u1_deltaS_gtheta}
\begin{aligned}
    \Delta S_A^{(2)} &\approx -\log \frac{\mathbb{E}_\mathbb{U} [{\rm tr}(\rho_{A,Q}^2)]}{\mathbb{E}_\mathbb{U} [{\rm tr}(\rho_{A}^2)]} \\
    &\approx -\log \frac{2^{|\bar{A}|} \exp\left[ 2 |A| \log^2\left|\tan\frac{\theta}{2}\right|\right] + 2^{|A|} \exp\left[ 2 |\bar{A}| \log^2\left|\tan\frac{\theta}{2}\right|\right] /\sqrt{\pi|A|} }{  2^{|\bar{A}|} \exp\left[ 2|A| \log^2\left|\tan\frac{\theta}{2}\right|\right] + 2^{|A|} \exp\left[ 2 |\bar{A}| \log^2\left|\tan\frac{\theta}{2}\right|\right]} \\
    &= -\log \frac{1 + 2^{|A|-|\bar{A}|} \exp\left[ -2 (|A|-|\bar{A}|) \log^2\left|\tan\frac{\theta}{2}\right|\right] /\sqrt{\pi|A|} }{  1 + 2^{|A|-|\bar{A}|} \exp\left[ -2 (|A|-|\bar{A}|) \log^2\left|\tan\frac{\theta}{2}\right|\right]} \\
    &= -\log \frac{1 + g(\theta)^{|A|-|\bar{A}|} /\sqrt{\pi|A|} }{  1 + g(\theta)^{|A|-|\bar{A}|}},
\end{aligned}
\end{equation}
where $|A|-|\bar{A}|=2|A|-N=2N(|A|/N-1/2)$ and
\begin{equation}
    g(\theta) = 2 \exp\left[ -2  \log^2\left|\tan\frac{\theta}{2}\right|\right].
\end{equation}
Note that $g(\theta)= 2$ for $\theta=\frac{\pi}{2}$ and $g(\theta)\geq 1$ for $\theta-\frac{\pi}{2}\in[-0.17\pi, 0.17\pi]$. Within this range, the expression above indicates that in the large size limit $N\rightarrow \infty$, $\Delta S_A^{(2)}\rightarrow 0$ if $|A|<N/2$ and $\Delta S_A^{(2)}\rightarrow \log\sqrt{\pi |A|}$ if $|A|>N/2$, quite similar to the result Eq.~\eqref{eq:asymm_no_symmetry} in the case of non-symmetric evolution except that the base factor $g(\theta)$ can deviate from $2$. The validity of Eq.~\eqref{eq:u1_deltaS_gtheta} can be verified by numerically computing the summations of binomial coefficients in Eqs.~\eqref{eq:u1_purity_rho_A} and \eqref{eq:u1_purity_rho_AQ}. As shown in Fig.~\textcolor{LinkColor}{2} of the main text, if $\theta=0.5\pi$, the results almost coincide with the non-symmetric case, consistent with the approximated analytical results Eq.~\eqref{eq:u1_deltaS_gtheta} and Eq.~\eqref{eq:asymm_no_symmetry}. If $\theta$ slightly deviates from $\frac{\pi}{2}$, e.g., $\theta=0.4\pi$, as shown in Fig.~\ref{fig:deltaS_A_04pi}, the overall magnitude of the curves is decreased but the main shape and trend remain qualitatively unchanged. 

\begin{figure}
    \centering
    \includegraphics[width=0.45\linewidth]{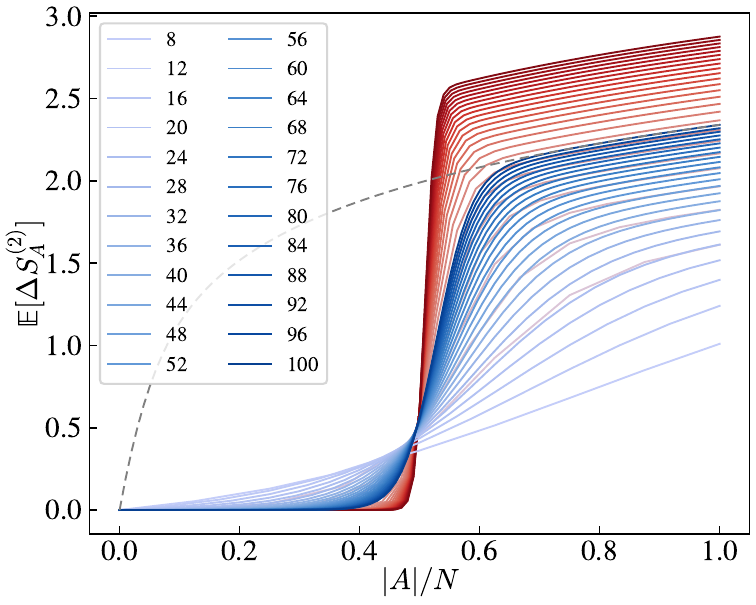}
    \caption{The average R\'enyi-2 entanglement asymmetry $\mathbb{E}[\Delta S^{(2)}_A]$ in the long-time limit of U(1)-symmetric random circuit evolution with the tilt angle $\theta=0.2\pi$ from the ferromagnetic initial state versus the subsystem size $|A|$. The numbers in the legend represent different system sizes $N$. The blue and red lines represent the results from U(1)-symmetric circuits and non-symmetric circuits, respectively. The grey dashed line represents the result for the tilted ferromagnetic initial state with $N=100$.}
    \label{fig:deltaS_A_02pi}
\end{figure}

\begin{figure}
    \centering
    \includegraphics[width=0.45\linewidth]{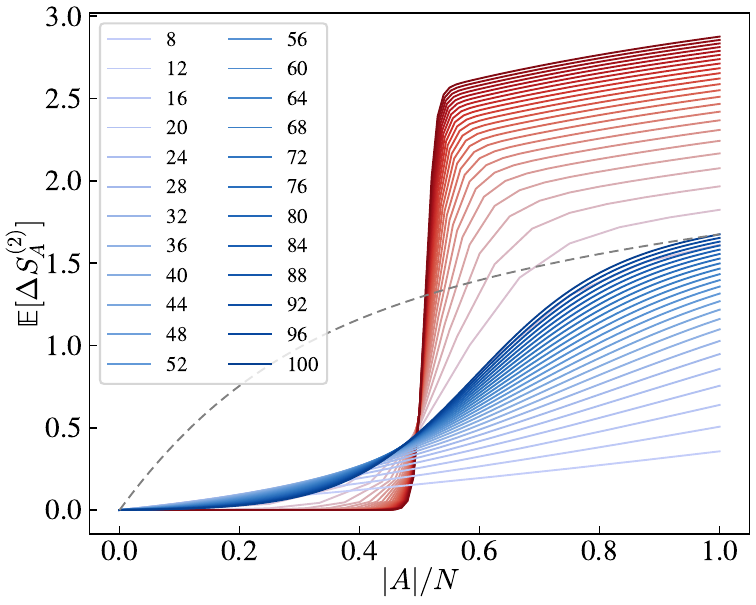}
    \caption{The average R\'enyi-2 entanglement asymmetry $\mathbb{E}[\Delta S^{(2)}_A]$ in the long-time limit of U(1)-symmetric random circuit evolution with the tilt angle $\theta=0.1\pi$ from the ferromagnetic initial state versus the subsystem size $|A|$. The numbers in the legend represent different system sizes $N$. The blue and red lines represent the results from U(1)-symmetric circuits and non-symmetric circuits, respectively. The grey dashed line represents the result for the tilted ferromagnetic initial state with $N=100$.}
    \label{fig:deltaS_A_01pi}
\end{figure}

A feature significantly different from the case of non-symmetric random circuits is that the average R\'enyi-2 entanglement asymmetry of the final state in the case of U(1)-symmetric random circuits is always lower than that of the initial state, regardless of the value of $|A|$, as shown by the blue curves in Fig.~\textcolor{LinkColor}{2} and Figs.~\ref{fig:deltaS_A_04pi}-\ref{fig:deltaS_A_01pi}. This is not necessarily true in the case of non-symmetric random circuits, as shown by the red curves in the corresponding figures, where the average entanglement asymmetry of the final state is the same for any initial state, allowing situations where the entanglement asymmetry of the final state is larger than that of the initial state. In this sense, non-symmetric random circuit evolution is featureless, whereas symmetric random circuit evolution holds a meaningful physical interpretation, i.e., it only leads to the restoration of the symmetry without further breaking the symmetry.

In addition, when $\theta$ slightly deviates from $\frac{\pi}{2}$, the factor $\log^2\left|\tan\frac{\theta}{2}\right|$ deviates from $0$ and hence $g(\theta)$ decreases from $2$. If $|A|<N/2$, then $g(\theta)^{|A|-|\bar{A}|}$ will increase and hence the late time  $ \mathbb{E} [\Delta S_A^{(2)}]$ will increase. On the other hand, for the initial tilted state, when $\theta$ decreases from $\frac{\pi}{2}$ to $0$, the U(1)-symmetry breaking is weakened, and the entanglement asymmetry $ \mathbb{E} [\Delta S_A^{(2)}]$ will decrease. The inverse order of $ \mathbb{E} [\Delta S_A^{(2)}]$ at the beginning and the end of the evolution suggests that there must exist cross points among the evolution curves with different $\theta$, consistent with the quantum Mpemba effect observed in the numerical experiments.

If $\theta$ deviates too much from $\frac{\pi}{2}$, or say $\theta$ is close to $0$, e.g., $\theta<0.33\pi$, then the Gaussian approximation above fails and there is no direct analytical evidence for the symmetry restoration at $|A|<N/2$. Direct numerical estimation using Eqs.~\eqref{eq:u1_purity_rho_A} and \eqref{eq:u1_purity_rho_AQ} shows that for small tilt angles such as $\theta<0.1\pi$, $ \mathbb{E} [\Delta S_A^{(2)}]$ will converge to a significant finite value in the long-time limit for large but finite system size $N$. As shown in Fig.~\textcolor{LinkColor}{2} of the main text, when $\theta=0.05\pi$, $ \mathbb{E} [\Delta S_A^{(2)}]$ increases with the system size for any $|A|$ up to $N=100$, which is very different from the large-$\theta$ case where $ \mathbb{E} [\Delta S_A^{(2)}]$ decreases with $N$ at $|A|<N/2$. This fact implies that if the symmetry breaking in the tilted ferromagnetic initial state is too weak, it will be hard to fully restore the symmetry by the U(1)-symmetric random circuit evolution, while those initial states with more severe symmetry breaking can successfully restore the symmetry by contrast. This is somehow an extreme case of the quantum Mpemba effect, i.e., instead of restoring slowly, the symmetry does not restore at all for initial states with small symmetry breaking. It is worth noting that the above discussion is restricted to the finite-size system with uniformly tilted ferromagnetic initial states.

Nevertheless, the increase of $ \mathbb{E} [\Delta S_A^{(2)}]$ with $N$ in the small-$\theta$ case will not last to the thermodynamic limit. This can be seen from Fig.~\ref{fig:deltaS_A_01pi} where $ \mathbb{E} [\Delta S_A^{(2)}]$ at $|A|<N/2$ increases with $N$ first and then decreases. This non-monotonic behavior will be more clear in Fig.~\textcolor{LinkColor}{3} of the main text, where we fix $|A|=N/4$ to see the curve of $ \mathbb{E} [\Delta S_A^{(2)}]$ versus $\theta$ for different $N$. We can see that for $0.2\pi\lesssim\theta\leq 0.5\pi$, $ \mathbb{E} [\Delta S_A^{(2)}]$ decreases monotonically with $N$ very fast down to zero while for $\theta\lesssim 0.2\pi$, $ \mathbb{E} [\Delta S_A^{(2)}]$ will first increase and then decreases with $N$, and converges to zero very slowly for a relatively large $N$. 

Importantly, one can see a prominent peak in Fig.~\textcolor{LinkColor}{3} of the main text, whose position is denoted as $\theta_{\mathrm{max}}$ below, gradually shifts towards $\theta=0$ as $N$ increases while the height of the peak is almost constant. Around this peak at $\theta_{\mathrm{max}}$, $ \mathbb{E} [\Delta S_A^{(2)}]$ takes significant finite values while is very close to zero away from this peak. The left end of this peak is just $\theta=0$. The right end of this peak can be properly defined as $\theta_c=2\theta_{\mathrm{max}}$. In other words, $\theta_c$ serves as a critical point where on the small-$\theta$ side the symmetry is not fully restored (persistent symmetry-breaking phase) while on the large-$\theta$ side, the symmetry is restored (symmetry-restored phase). A schematic finite-size crossover phase diagram is shown in Fig.~\textcolor{LinkColor}{3} of the main text.

However, the ``symmetry breaking phase'' is not a true phase in the sense of the thermodynamic limit because $\theta_c$, albeit slowly, scales with the system size $N$ in the scaling of $1/\sqrt{N}$, as shown in the inset of Fig.~\textcolor{LinkColor}{3}. That is to say, in the thermodynamic limit, regardless of the value of the tilt angle $\theta$, the subsystem symmetry will always be restored for $|A|<N/2$ which is consistent with the eigenstate thermalization hypothesis. But conversely, for any finite-size system, there always exists a ``critical point" $\theta_c \approx 1.13 \pi/\sqrt{N}$, such that the subsystem symmetry cannot be restored for tilted ferromagnetic initial state with $\theta<\theta_c$. 

It is worth mentioning that this lack of symmetry restoration is only observed in slightly tilted ferromagnetic initial states. As a counterexample, for tilted N\'eel states, the subsystem symmetry will always be restored by the U(1)-symmetric random circuit at late time as long as $|A|<N/2$ regardless of the tilt angle. This can be attributed to the fact that the N\'eel state is in the largest symmetry sector of exponentially large dimensions, while the ferromagnetic state is in a small symmetry sector of one dimension in the $\theta=0$ limit.

\end{document}